\documentclass[twocolumn]{aastex6} 
\usepackage{bm}
\usepackage{amsmath,amssymb}
\usepackage{ulem}
\def\BIDISPERSE{\iffalse}
\def\CONCLUSION{\iffalse}
\def\APPENDIX{\iffalse}

\newcommand\aveepsilon{\langle\varepsilon\rangle}

\begin{document}


\title{Dust coagulation regulated by turbulent clustering in protoplanetary disks}
\shorttitle{Dust coagulation regulated by turbulent clustering}




\author{Takashi Ishihara\altaffilmark{1},Naoki Kobayashi\altaffilmark{2},Kei Enohata\altaffilmark{2},Masayuki Umemura\altaffilmark{3}, Kenji Shiraishi\altaffilmark{4}}



\altaffiltext{1}{Graduate School of Environmental and Life Science, Okayama University, Okayama 700-8530, Japan; ishihara@ems.okayama-u.ac.jp}
\altaffiltext{2}{Department of Computational Science and Engineering, Graduate School of Engineering, Nagoya University, Nagoya 464-8603, Japan}
\altaffiltext{3}{Center for Computational Sciences, University of Tsukuba, Tsukuba 305-8577, Japan; umemura@ccs.tsukuba.ac.jp}
\altaffiltext{4}{Institute of Materials and Systems for Sustainability, Nagoya University, Nagoya 464-8601, Japan}

\begin{abstract}

The coagulation of dust particles is a key process in planetesimal formation. 
However, the radial drift and bouncing barriers are not completely resolved, 
especially for silicate dust. 
Since the collision velocities of dust particles are 
regulated by turbulence in a protoplanetary disk, the turbulent clustering should 
be properly treated. To that end, direct numerical simulations (DNSs) of 
the Navier Stokes equations are requisite. In a series of papers, Pan \& Padoan 
used a DNS with the Reynolds number $Re \sim 1000$. 
Here, we perform DNSs with up to $Re = 16100$, which allow us to track the motion 
of particles with Stokes numbers of $0.01\lesssim St \lesssim 0.2$ in the inertial range. 
By the DNSs, we 
 confirm that the rms relative velocity of particle pairs is smaller 
by more than a factor of two, compared to those by Ormel \& Cuzzi (2007). 
The distributions of the radial relative velocities are highly non-Gaussian. 
The results are almost consistent with those by Pan \& Padoan or Pan et al. at low-$Re$. 
Also, we find that the sticking rates for equal-sized particles are much higher than 
those for different-sized particles. 
Even in the strong-turbulence case with ƒ$\alpha$-viscosity of $10^{-2}$, 
the sticking rates are as high as $\gtrsim 50\%$ 
and the bouncing probabilities are as low as $\sim 10\%$ for equal-sized particles of $St \lesssim 0.01$. 
Thus, the turbulent clustering plays a significant role for the growth of cm-sized compact aggregates (pebbles) 
and also enhances the solid abundance, which may lead to the streaming instability in a disk.

\end{abstract}


\section{Introduction} \label{sec:intro}

Planetesimals are thought to be the precursors of both earth-like planets as well as 
the cores of gas giants and ice giants. 
The formation of planetesimals has been a longstanding and perplexing obstacle toward
the full understanding of the origins of planets. 
Planetesimals are widely believed to form as a consequence of the hierarchical coagulation from 
submicron-size dust particles to kilometer-size bodies in protoplanetary disks
\citep[e.g.][]{Lissauer1993,Chiang2010,Johansen2014}.
The growth, however, faces several difficulties, such as bouncing, fragmentation, and radial drift barriers.
As the gas disk is partially pressure-supported in the radial direction, 
the gas rotates with a sub-Keplerian velocity.
The resultant friction by the gas forces dust particles to drift toward the central star. 
In particular, the centimeter- to meter-size particles have large drift velocities and may rapidly fall into 
the central star on a timescale of a few 100 yr 
\citep{Adachi1976,Weidenschilling1977},
which is often referred to as the meter-size problem.
Therefore, the growth timescale is required to be shorter than the drift timescale
for the successful formation of planetesimals. 
However, as the dust particles grow, they become less sticky, and the high-velocity collisions may
lead to bouncing or fragmentation instead of coagulation 
\citep{Blum2008,Brauer2008a,Brauer2008b,Birnstiel2009,Birnstiel2012,Gttler2010,Zsom2010,Zsom2011,Windmark2012}.

The streaming instability is a potential mechanism to circumvent the radial drift barriers
\citep{YG05}. In order for this instability to work successfully, the formation of
cm-sized compact aggregates (pebbles) and the enhancement of solid abundance 
in a protoplanetary disc are crucial
\citep{Johansen2009,BS2010,Johansen2014,Carrera2015,Ida2016,Yang2017}.
On the other hand, in recent years, $N$-body molecular-dynamics simulations have shown 
that {\it fluffy} dust aggregates have the potential to overcome these bouncing or fragmentation barriers
\citep{Wada2009,Paszun2009,Wada2013,Meru2013,Seizinger2013,Gunkelmann2016}.
\cite{Wada2013} derived the critical collision velocity $u_{\rm c}$, below which 
fluffy dust aggregates can coalesce: $u_{\rm c}\simeq 60-80$ m s$^{-1}$ for ice dust
and $u_{\rm c}\simeq 6-8$ m s$^{-1}$ for silicate dust.
This implies that icy aggregates overcome the fragmentation barriers more easily, compared to
silicate aggregates.

Protoplanetary disks inevitably become turbulent due to the differential rotation, and therefore
the collision velocities are regulated by the turbulent motion in a wide range of particle sizes, 
from 1 mm to 10 m 
\citep[e.g., a review by ][]{Johansen2014}.
\cite{Volk1980} 
built up a framework set (the V\"olk-type model) based on a Langevin approach
for the nonlinear response of dust particles to turbulent eddy motion, 
which was further developed by 
\cite{Markiewicz1991}.
\cite{Ormel2007}(OC07) 
provided closed-form analytical approximations 
to the V\"olk-type model. 
\cite{Okuzumi2012},
based on an analytic formula by OC07,
have simulated the growth of fluffy icy aggregates outside the snow line,
taking into account the change in aggregate porosities, and found
that the porosity evolution enables the icy aggregates to grow across the radial drift barrier. 
\cite{Kataoka2013} 
have explored the compression of fluffy aggregates
and shown that the aggregates can evolve into compact icy planetesimals.
The abovementioned works revealed that icy planetesimals can form in a wide range beyond the snowline 
in protoplanetary disks. 
However, the difficulties for silicate planetesimal formation are difficult to alleviate,
because the critical collision velocity for silicate dust is smaller by an order of magnitude 
than that for icy dust. 

The actual sticking rates of dust particles are strongly dependent on the probability 
distribution function (PDF) of the collision velocities. 
Even if the root mean square (rms) collision velocity exceeds the critical value, 
a subset of the dust particles can have collision velocities lower than the critical one
and eventually grow, evading the fragmentation barrier. 
Employing the rms collision velocity derived by OC07,
\cite{Windmark2012} and \cite{Garaud2013} 
have explored the effects of the PDF on the collisional dust growth barriers, under the assumption of a Gaussian (Maxwellian) distribution. 
However, the PDFs of turbulence-induced relative velocities are found to be highly 
non-Gaussian by numerical, experimental, and theoretical studies 
\citep{Sundaram1997,Gustavsson2008,Gustavsson2011,Hubbard2012}.
In addition, it is known that particles with small inertia preferentially concentrate in low-vorticity, 
high-strain regions during turbulence due to the centrifugal mechanism of the vorticity 
\citep{Maxey1987,Squires1991,Fessler1994}.
Furthermore, effects such as ``caustics'' \citep{Wilkinson2006} and the ``sling effect'' \citep{Falkovich2007}
allow the particles with large inertia to become less coupled with the local fluid velocity field and 
assemble from different regions \citep[e.g., see][]{Bragg2014}. 
Such ``turbulent clustering'' effects are significant when considering the process of the collisional coagulation 
of dust particles in protoplanetary disks.

To treat the turbulent clustering properly, the direct numerical simulation (DNS) of 
the Navier-Stokes equations coupled with tracking dust particle motions are requisite. 
In the DNS, the smallest eddies in the turbulence are resolved without 
introducing numerical viscosity and turbulence models. 
\cite{Pan2011} 
handled the collision statistics with the turbulent clustering, using an Eulerian formulation 
instead of the Navier-Stokes equations. Then, in a series of the papers by 
\cite{Pan2013}(PP13), \cite{Pan2014_II,Pan2014_III,Pan2014_IV}, and \cite{Pan2015}(PP15), 
they used a DNS of the Navier-Stokes equations in the context of planetesimal formation.
By analyzing the DNS data, they studied the statistics of colliding dust grains including the radial relative velocity, its probability distribution and the collision rate between dust grains. However, 
Reynolds number dependence of the results has not been investigated yet.

The motion of particles in turbulence is characterized by the Stokes number given by $St=\Omega \tau_{\rm p}$,
where $\Omega$ is the Keplerian frequency at a radial distance and $\tau_{\rm p}$ is
the stopping time by the gas friction. In addition, another Stokes number can be defined by the turnover timescale $\tau_\eta$
of the smallest eddies of turbulence as $St_\eta=\tau_{\rm p}/\tau_\eta$. 
The $St_\eta$ is determined by the resolution of the simulations, and has a relation 
$St_\eta/St \propto Re^{1/2}$, where $Re$ is the Reynolds number.
According to Kolmogorov theory, as the Reynolds number increases, 
the inertial range of turbulence that regulates the particle dynamics becomes wider. 
The scale ratio between the largest and the smallest eddies is known to increase in proportion 
to $Re^{3/4}$. 

Considering the molecular viscosity of a protoplanetary disk,
the Reynolds number is estimated to be as high as $Re=10^{10}(\alpha/10^{-2})(R/{\rm AU})^{-3/2}$
in the Minimum-Mass Solar Nebula (MMSN) Model \citep{Hayashi1981}, 
where $\alpha$ is the turbulence parameter and $R$ is the radial distance from the central star.
However, even if the Reynolds number is smaller than this value, we can trace
the particle behavior over a range of Stokes numbers according to the simulated inertial range of turbulence.
If we focus on particle sizes from millimeters to meters, 
simulations of $Re>O(10^4)$ are required. In the simulations by PP13, 
the resolution was $Re \simeq 10^3$, which corresponds to $St_\eta/St=23.5$ and realizes
the inertial range of turbulence over only one order of magnitude in linear dimensions.
However, a recent development in supercomputers allows us to perform 
particle tracking simulations based on the DNS 
at Reynolds numbers as high as $Re>O(10^4)$ \citep{Ishihara2015}

In this paper, we perform high Reynolds number DNSs, where the number of grid points and the Reynolds number
are up to $N^3=2048^3$ and $Re =16100$, respectively,
which corresponds to $St_\eta/St=85$ and can realize
the inertial range over two order of magnitude in linear dimensions.
By these DNSs, we obtain the rms relative velocity for particle pairs and 
the PDF of the collision velocities. 
We find that the rms relative velocity is smaller than that 
derived by the V\"olk-type model developed by OC07. 
In addition, we discuss the growth timescale of dust aggregates and 
the sticking rates in the context of overcoming the drift and fragmentation barriers.

In Section 2, we present the method of our particle tracking simulation, 
based on the DNS of forced incompressible homogeneous isotropic turbulence. 
The statistics of the motion of the particles obtained by the DNS are shown in Section 3. 
The statistics include the rms relative velocity, the collision kernel, 
and the PDF of the radial relative velocities. The results are compared 
to the V\"olk-type model by OC07. 
In Section 4, assuming the MMSN model, we assess the collision timescale 
for both compact and fluffy aggregates and the sticking rates 
of dust particles. 
Our conclusions are summarized in Section 5.

\section{Particle Tracking using DNS of turbulence} 

\subsection{DNS of forced incompressible turbulence}

\begin{figure*}[t]
\begin{center}
\includegraphics[scale=.60]{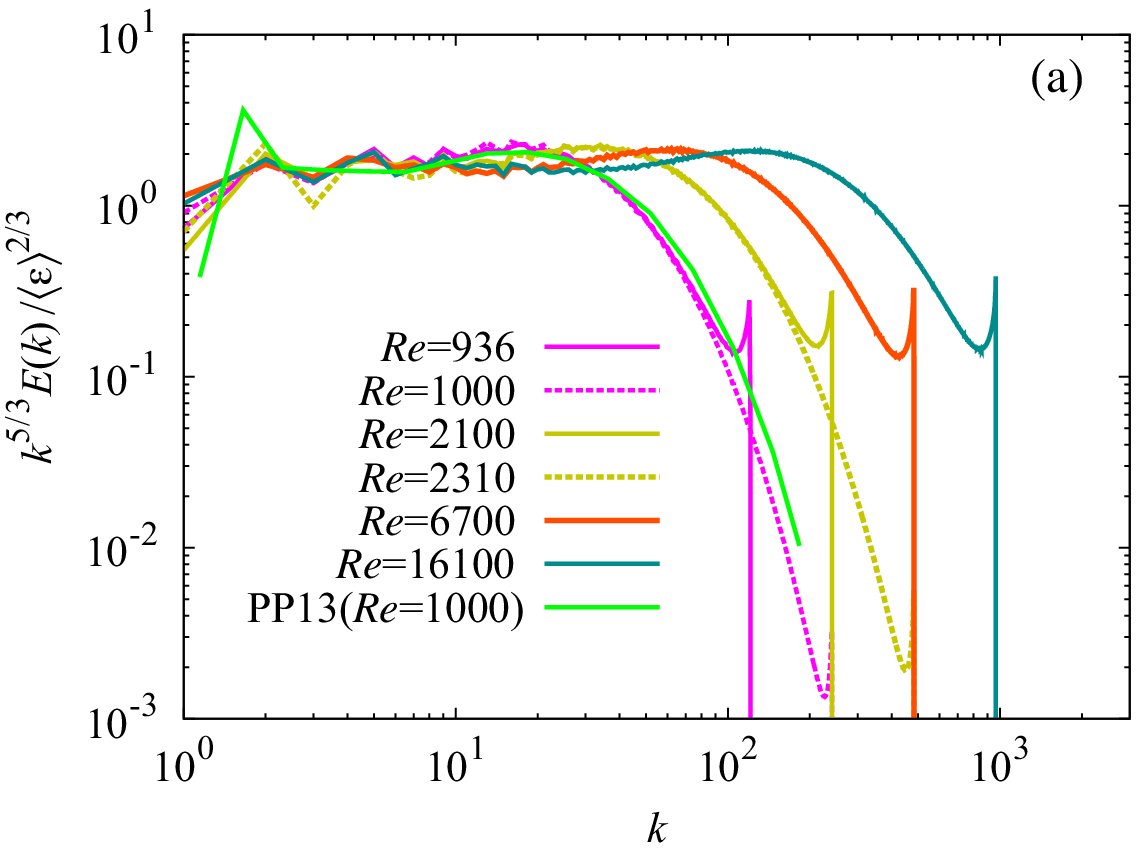}\quad\quad
\includegraphics[scale=.60]{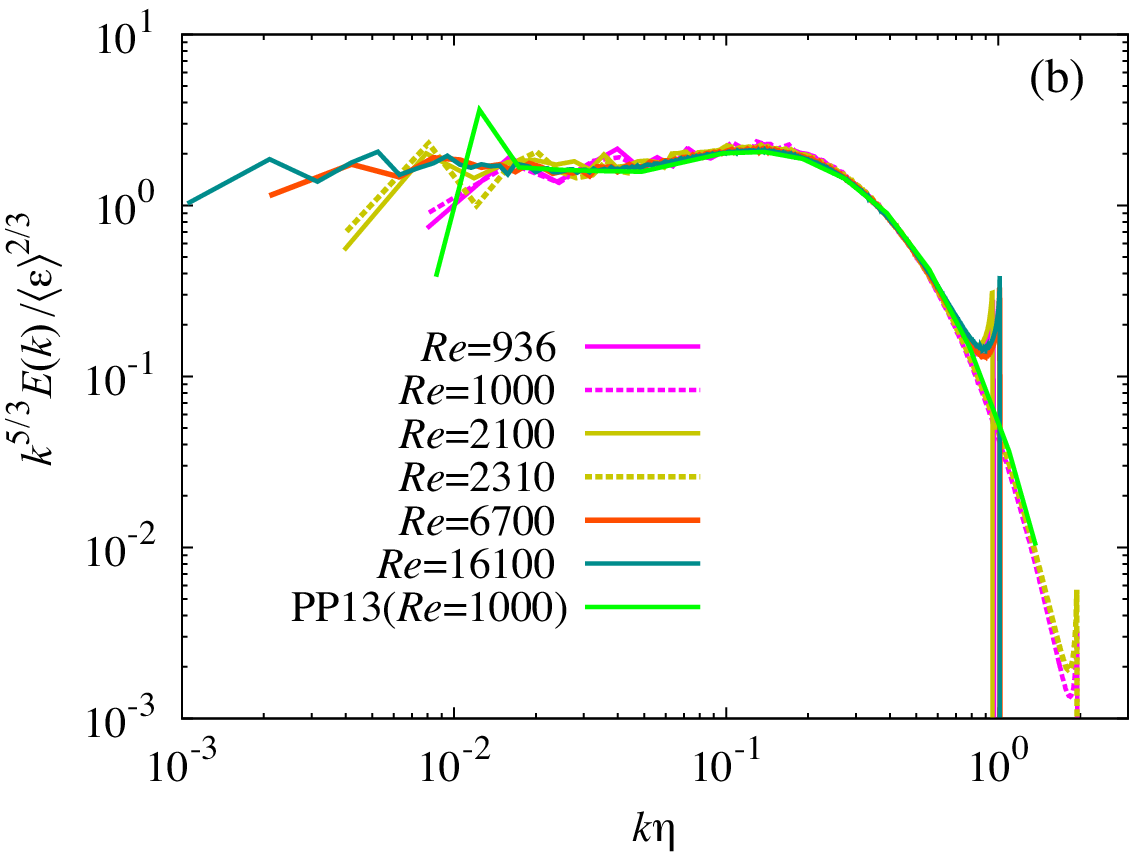}
\end{center}
\caption{Compensated energy spectra of the turbulence, $k^{5/3}E(k)/\langle\varepsilon\rangle^{2/3}$, 
for different Reynolds numbers $Re$,
(a) as a function of $k$ and (b) as a function of $k\eta$. The data from \cite{Pan2013}(PP13) are also plotted for comparison.
\label{fig:sp}}
\end{figure*}

In protoplanetary disks, the turbulence is known to be subsonic and 
thus essentially incompressible \citep{Hayashi1981}. 
Therefore, in this paper, we consider three-dimensional turbulence of an incompressible fluid of unit density
that obeys the Navier-Stokes (NS) equations
\begin{equation} 
\label{NS}
 \frac{\partial {\bm u}}{\partial t} + \left(\bm u \cdot \nabla \right) \bm u 
 = -\nabla p + \nu \nabla ^2 \bm u + \bm f 
\end{equation}
and the continuity equation
\begin{equation} 
\label{NS divu}
 \nabla \cdot \bm u = 0,
\end{equation}
where 
${\bm u}$, ${p}$, ${\nu}$, and ${\bm f}$ are the velocity, pressure, kinematic viscosity, and external force, respectively.
The numerical method used in the DNS is essentially identical to that used in \cite{Yokokawa2002}
and \cite{Kaneda2003}, which is briefly reviewed here for the convenience.
(The readers may refer to \cite{Yokokawa2002} and \cite{Morishita2015} for the details of parallel computations.)
In the DNS, the Navier-Stokes equations are solved by a fully alias-free Fourier spectral method, 
where aliasing errors are removed by the so-called phase-shift method.
(Note that the Fourier spectral method gives the spatial derivatives to spectral accuracy and is not subject to the numerical viscosity. 
On the other hand, in finite difference schemes, the spatial derivatives are calculated 
to finite accuracy and the numerical viscosity is inevitably included.)
The computation domain is assumed to be $2\pi$-periodic in each Cartesian coordinate direction, so that both the minimum wavenumber and the wavenumber increment in the DNS are unity.
The maximum wavenumber is given by $k_{\rm max}=\sqrt{2}N/3$, where $N$ is the number of grid points in each of the Cartesian coordinates in real space.
Time integration is achieved using a fourth-order Runge-Kutta method with a constant time increment. 

The forcing ${\bm f}$ that generates turbulence is given by $\hat{\bm f}({\bm k})=-c\hat{\bm u}({\bm k})$ in the wavevector space, 
where $\hat{\bm f}$ and $\hat{\bm u}$ are Fourier transforms of ${\bm f}$ and ${\bm u}$, respectively.
The value of $c$ is set as non-zero only in the wavenumber range of $k<2.5$ and 
is adjusted at every time step so as to keep the total kinematic energy, $E \equiv 3 u'^2/2=U^2/2$, 
almost time-independent $(\approx 0.5)$.
Here, $u'$ is the root mean square (rms) value of the fluctuating velocity in one direction and $U=\left<{\bm u}\cdot{\bm u}\right>^{1/2}$ is the three-dimensional (3D) rms  of flow velocity ${\bm u}$. 
A similar forcing was used in \cite{Kerr1985}, \cite{Vincent1991}, and \cite{Jimenez1993}.
The integral length scale $L$ and the eddy turnover time $T$ are defined by 
$L=\pi/(2U^2)\int_0^k E(k)/k dk$ and $T=L/U$, respectively. 
$T$ corresponds to $\Omega^{-1}$ in the actual dimensions.
The variations of $L$ and $T$ in steady turbulence generated by the forcing
are within 10\% of the mean \citep[][]{Ishihara2007}. 

\begin{table*}[t]
\caption{Simulation (DNS) parameters and turbulence characteristics.  $Re=u'L/\nu$, $T=L/u'$ and $\tau_\eta\equiv (\nu/\aveepsilon)^{1/2}$.}
\label{tabDNS}
\begin{center}
\begin{tabular}{cccccccccc} \hline\hline
Run & $N^3$ & $Re$ & $k_{\rm max}$ & $\Delta t(\times 10^{-3})$ & $\nu(\times 10^{-4})$ & $L$& $\eta(\times 10^{-3})$& $T$ & $\tau_{\eta}$ \\ \hline
256-1 & $256^3$ & 936 & 121    & 1.0            & 7.0        &1.13  & 7.97    &1.96    & 0.091   \\
 512-1 & $512^3$ & 2100 & 241   & 1.0            & 2.8        &1.02 & 3.95       &1.77  & 0.056   \\
1024-1 & $1024^3$ & 6710 & 471    & 0.625           & 1.1       &1.28  & 2.10     &2.21  & 0.040   \\
2048-1 & $2048^3$ & 16100 & 732    & 0.4            & 0.44       &1.23  & 1.05     &2.13  & 0.025   \\ \hline
 512-2 & $512^3$ & 1000 & 241    & 1.0            & 7.0         &1.10 & 8.10    &2.21    & 0.094   \\
1024-2 & $1024^3$ & 2310 & 483    & 0.625           & 2.8        &1.21 & 4.03   &1.94      & 0.058   \\
\hline
\end{tabular}
\end{center}
\end{table*}

We use the velocity fields obtained by the DNS in \cite{Kaneda2003} and \cite{Ishihara2007} 
as the initial conditions for the present study and adopt the same values of the kinematic viscosity. 
Therefore, each velocity field at $t=0$ is in a statistically steady state of turbulence and 
we do not have any initial transient period in the turbulence field used 
in the particle tracking simulation. The details of the turbulence characteristics 
of the DNS data are found in \cite{Ishihara2007}.

In Table \ref{tabDNS}, we show a summary of the DNS parameters and turbulence characteristics. 
In the DNSs, the value of $\nu$ is chosen so that $k_{\rm max}\eta=1$ or $2$, 
where $\eta$ is the Kolmogorov micro length scale given by 
$\eta=(\nu^3/\left<\varepsilon\right>)^{1/4}$ with the mean energy dissipation rate $\left<\varepsilon\right>$.
The value of $k_{\rm max}\eta$ represents a small-scale resolution.
Each ``Run'' in Table \ref{tabDNS} is named by the combination of the values of $N$ and $k_{\rm max}\eta$. 
It is known that the sensitivity/insensitivity to the values of $k_{\rm max}\eta$ 
depends on the statistical quantity to be studied \citep{Yamazaki2002,Watanabe2007,Schumacher2007,Donzis_etal2008,Yeung2015}.
It will be shown that the collision statistics are not so sensitive to 
the $k_{\rm max}\eta$ values, provided that $k_{\rm max}\eta \gtrsim 1$.

Figure \ref{fig:sp} shows the compensated energy spectra, 
$k^{5/3}E(k)/\langle\varepsilon\rangle^{2/3}$, 
of the generated turbulence for different values of $Re$ and $k_{\rm max}\eta$.
The spectrum of the velocity field used in PP13 and PP15is also shown for comparison.
The horizontal range corresponds to the ``inertial range", where $E(k)\propto k^{-5/3}$.
In Figure \ref{fig:sp}, we observe that the horizontal range becomes wider with increasing $Re$.
Therefore, it is expected that the inertial range may be satisfactorily resolved by our DNS.
In the DNS of $Re=16100$, we realize $L/\eta=1.2\times 10^3$ and $T/\tau_\eta=85$, respectively, 
which are much larger than $L/\eta=1.4\times 10^2$ and $T/\tau_\eta=23.5$ for $Re=1000$.

We recognize that the compensated energy spectra obtained by our DNSs have a pile-up near $k=k_{\max}$. 
It is known that such a pile-up 
is caused by the wavenumber truncation in the DNS based 
on a Fourier spectral method. It is also known that such a pile-up 
does not appear in the turbulence simulation using finite difference methods. 
Therefore, the difference between our spectra and the spectrum of PP13 and PP15 
near $k=k_{\max}$ comes from the difference in the numerical methods.

To see an effect of the wavenumber truncation in the Fourier spectral method, we compare 
the energy spectra between runs 512-1 and 1024-2 (and also between runs 256-1 and 512-2) in Figure \ref{fig:sp}. 
They are slightly different from each other 
in the low wavenumber range ($k<3$) and in the high wavenumber range ($k\eta>0.7$).  
The difference in the low wavenumber range is presumably caused by the difference in the energy-containing eddies at the forcing scales, while the difference at high wavenumbers is caused by the wavenumber truncation.
Note that in contrast to the high and low wavenumber ranges, the difference between the spectra in the intermediate range for the two runs is very small. 
This comparison suggests that the collision statistics are not sensitive to 
the difference between $k_{\rm max}\eta\sim 1$ and $k_{\rm max}\eta\sim 2$, 
if they are insensitive to the detail of the fine-scale statistics in the energy dissipation range of the turbulence.
In this study, we will mainly present the results obtained by the DNS data with $k_{\rm max}\eta\sim 1$.
However, we will also show the results of the DNS data with $k_{\rm max}\eta\sim 2$ to confirm the insensitivity of the quantity to the small-scale resolution (see figures \ref{relative_velocities2} and \ref{fig:collisionkernel}).

\subsection{Particle tracking simulation}

We consider the motion of small solid particles with density $\rho_{\rm s}$ in a gas flow with density $\rho_{\rm g}$.
The ratio $\beta=\rho_{\rm s}/ \rho_{\rm g}$ is assumed to be much larger than unity.
Then, the equation of motion of each inertial particle is given by
\begin{equation} 
 \frac{d \bm X}{dt} = \bm V, \quad
 \frac{d \bm V}{dt} = \frac{1}{\tau_{\rm p}}(\bm u - \bm V),
\label{eq:p}
\end{equation}
where $\bm X$, $\bm V$, and ${\tau_{\rm p}}$ are the position, velocity, and stopping time of the particle, respectively, and
${\bm u}$ is the velocity of the fluid flow at $\bm X$ \citep[see, e.g.,][]{Davila2001}.
Equation (\ref{eq:p}) is solved with the fourth-order Runge-Kutta method, where the time step ($\Delta t$) is set to be twice as large as that used to solve Equation (\ref{NS}) (because we need ${\bm u}({\bm x},n\Delta t/2)$ in this scheme).
The velocity ${\bm u}$ at the particle position is evaluated by using an interpolation method.
To obtain accurate statistics, we use the cubic spline interpolation \citep[see][]{Yeung1988}.
The interpolation scheme is implemented by solving tridiagonal matrix problems in parallel with the method developed by \cite{Mattor1995}. See \cite{Ishihara2015} for the actual parallel implementation.

In this paper, we normalize Equation (\ref{eq:p}) in two ways.
One is by using $L$ and $T$ and the other is by using $\eta$ and $\tau_\eta\equiv (\nu/\aveepsilon)^{1/2}$.
In the former, the motion of the particles is characterized by the Stokes number given 
by $St=\tau_{\rm p}/T$, and in the latter, the motion is characterized by another Stokes number, given by $St_\eta=\tau_{\rm p}/\tau_\eta$.
Note that $St_\eta/St=T/\tau_\eta\propto Re^{1/2}$. In the case of the typical protoplanetary disk of $Re=O(10^{10})$, the ratio is $O(10^5)$. 
However, even if the Reynolds number is smaller than this value, we can trace
the particle behavior over a range of Stokes numbers according to the simulated inertial range of the turbulence.
In our largest DNS of $Re=16100$, the value of $St_\eta/St$ is $85$. 

In our particle tracking simulation, we set seven different values of $St$ as follows.
\begin{eqnarray} 
St&=& 0.00158,\ 0.00632,\ 0.0158, 0.0316, \nonumber \\ & & \ 0.0632, \ \ 0.158, \ 0.316,\ \ \hbox{for}\ N=512, \\
St&=& 0.00188,\ 0.0038,\ \ 0.0094, 0.0188, \nonumber \\ & & \ 0.038,\ 0.094,\ 0.188, \ \ \hbox{for}\ N=1024, \\
St&=& 0.00117,\ 0.00587, \ \ 0.0117, 0.0234, \nonumber \\ & & \ 0.0587, \ 0.117, \ 0.234,\ \ \hbox{for}\ N=2048.
\end{eqnarray}
The corresponding values of $St_\eta=St (T/\tau_\eta)$ are 0.1, 0.2, 0.5, 1, 2, 5, and 10 for $N=512,\ 1024$ and
0.1, 0.5, 1, 2, 5, 10, and 20 for $N=2048$. 

For each value of $St$, we maximally track $512^3$ particles. The particles are distributed randomly in the whole computational domain at $t=0$ with initial velocity ${\bm V}_0={\bm 0}$. The statistics related to the particle motions are taken at $t=3T$ in the following analyses. 
By performing several preliminary runs, we have confirmed that the particle statistics 
become almost time-independent after $t=3T$.

In this paper, we focus on the particles with the Stokes number less than $0.3$. The collision statistics of particles with $St=O(0.1)$ 
 are expected to be mainly affected by the eddies in the inertial range which are able to simulate properly by high-resolution DNSs of homogeneous isotropic turbulence using appropriate forcing methods. Therefore, the aim of this paper is to provide reliable results of collision statistics of the particles in the inertial range by performing a series of the high-resolution DNSs. Since the relative velocity attains its maximum at $St\sim 1$, such collisions may be important to discuss the fragmentation barrier. However, it is also true that such collisions occur via the processes of collision and sticking of the dusts with smaller values of $St$ which are dominated by eddies in the inertial range.  Therefore, we took a bottom-up approach.

\section{Numerical results}

\subsection{Relative velocity}

\begin{figure}[t]
\begin{center}
\includegraphics[scale=.65]{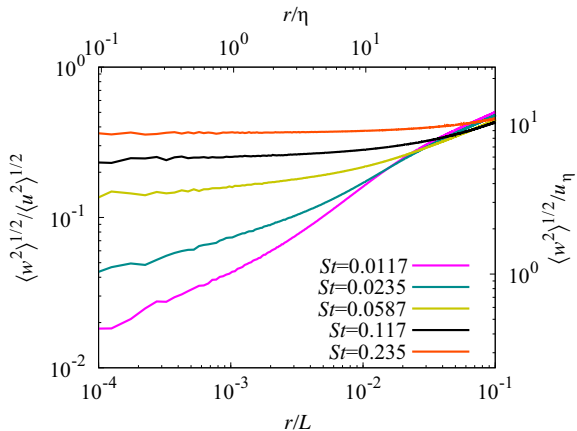}
\end{center}
\caption{rms relative velocity as a function of the separation ($r$) between a pair of particles.
The data are plotted for each $St$ number for $512^3$ particles at $t=3T$ 
in the DNS of Run2048-1 ($Re=16100$) and normalized by 
$\langle u^2\rangle^{1/2}\equiv\langle {\bm u}\cdot{\bm u}\rangle^{1/2}$ or $u_\eta$.
\label{relative_velocities1}}
\end{figure}

The relative velocity between a pair of particles 
with a separation $r$ and its rms value are calculated at locations ${\bm X}_1$ and ${\bm X}_2$ as 
${\bm w}={\bm V}_1-{\bm V}_2$ and
\begin{equation}
\langle w^2 \rangle^{1/2}=\big\{\frac{1}{N_{\rm p}}\sum_{|{\bm X}_1-{\bm X}_2|=r}{\bm w}^2\big\}^{1/2},
\end{equation}
respectively, where ${\bm V}_1$ and ${\bm V}_2$ are the velocities of the particles 
and $N_{\rm p}$ is the number of pairs.
Figure \ref{relative_velocities1} shows the $r$-dependence of the rms relative velocity between equal-sized particles.
Actually, collision velocities should be measured at particle sizes, 
which are in general much smaller than the Kolmogorov length scale in turbulence in protoplanetary disks. 
To estimate the rms relative velocity at a separation of less than $\eta/4$ in the DNSs, 
one may need a much larger number of particles than that used in our DNSs. 
However, Figure \ref{relative_velocities1} demonstrates that 
the relative velocity for the particles of $St>0.05$ is not significantly affected 
by eddies smaller than $\eta(\sim 10^{-3}L)$
and therefore almost constant at a separation $r \lesssim 10^{-3}L$.

\begin{figure}[t]
\begin{center}
\includegraphics[scale=.6]{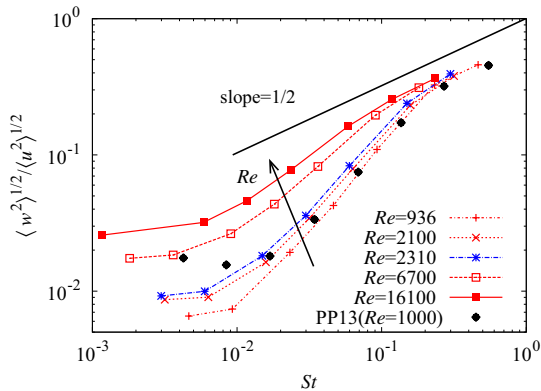}
\end{center}
\caption{rms relative velocity at $r=10^{-3}L$ as a function of the $St$ number.
The dependence on the $Re$ numbers are shown for 
$Re=936$ (dot-dashed line), $Re=2100$ (dotted line), $Re=2310$ (chain line), $Re=6700$ (dashed line), and
$Re=16100$ (solid line). 
The data from PP13 ($Re=1000$, $r=\eta/4\sim 2\times 10^{-3}L$) are also plotted for comparison.
\label{relative_velocities2}}
\end{figure}

Figure \ref{relative_velocities2} presents the $St$-dependence of the rms relative velocity 
at a fixed small separation of $r=10^{-3}L$ for different Reynolds numbers.
Here, $r/L=10^{-3}$ approximately corresponds to 
$r/\eta=1/8, 1/4, 1/2,$ and $1$ for runs at $Re=936, 2100$ (and $2310$), $6700$, and $16100$, respectively.
(The other comparison for different $Re$ values using a fixed separation of $r/\eta$ would be also possible.
However, here we are interested in $St(=\tau_{\rm p}/T)$ dependence of the rms relative velocity.
So, we measure the relative velocity 
using a fixed small separation $r$ normalized by $L$.)
The data from PP13 ($Re=1000$, $r=\eta/4$) agree well with our data ($Re=936$, $r=\eta/8$) in the range of large $St$ $(\gtrsim 0.1)$. 
They deviate from our results in the range $St \lesssim 10^{-2}$.
This deviation is presumably due to the difference in the value of $r/\eta$,
and comes from the range of $St_\eta<1$. However, in this paper,
we focus on the collision statistics in the range $St_\eta>1$.
In addition, the data for $Re=2100$($k_{\rm max}\eta=1$) agree well with the data for $Re=2300$($k_{\rm max}\eta=2$).
This agreement indicates that the relative velocities at small separations are not so sensitive to the difference in the value of $k_{\rm max}\eta$.

In Figure \ref{relative_velocities2}, we perceive that the relative velocity 
at the small separation of $r/L=10^{-3}$ is an increasing function of $Re$ for a fixed value of $St$. 
Also, we observe that the curves tend to approach a line with slope $1/2$ 
at the portion of larger values of $St (\gtrsim 0.1)$ for $Re>10^4$. 
This dependence ($\langle w^2\rangle^{1/2}\propto St^{1/2}$) is consistent 
with the inertial range scaling of the relative velocity in V\"olk-type models; see PP15.

\subsection{Bidisperse case}

\begin{figure}[t]
\begin{center}
\includegraphics[scale=.6]{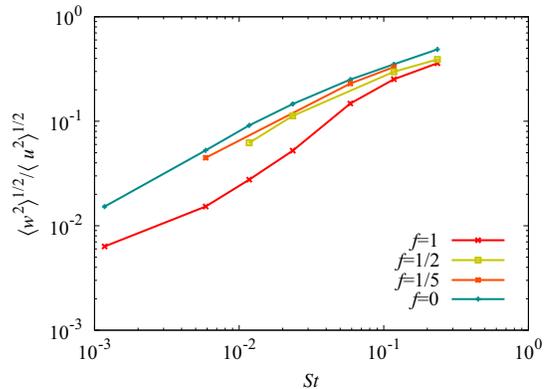}
\caption{rms relative velocity at $r = \eta/4$ for particle pairs with fixed Stokes ratios, $f\equiv St_2/St_1=1$, 
$1/2$, $1/5$, and $0$, where $f=1$ corresponds to the monodisperse case 
and $f=0$ to the relative velocities against the turbulent flow itself. 
The data are obtained at $t=3T$ in Run 2028-1 ($Re=16100$). 
 \label{fig:bid1}}
 \end{center}
\end{figure}

In the above, we considered the monodisperse case of the identical particles. 
Here, we present the results for bidisperse cases, i.e., the relative velocities of particles 
with different $St$ numbers ($St_1$ and $St_2$).
Figure \ref{fig:bid1} shows the $St(=St_1)$-dependence of the rms relative velocity 
between the particles with different ratios of $f\equiv St_2/St_1$.
To compare the relative velocities at a scale of smallest eddies, we show the results
at $r = \eta/4$.
We find that the rms relative velocity is higher for the different-sized particles than for the equal-sized particles. 
This trend is prominent for $St \approx 10^{-2}$.
The result can be understood by considering that the equal-sized particles with small separations
move in the same way in a turbulent flow, and therefore the relative velocities remain low.

It is known that collisions between particles at high speeds may lead to bouncing or fragmentation.
Therefore, the sticking rate of collision particles depends on the statistics of relative velocities.
The present results suggest that the collisions between the identical particles are 
more appropriate for sticking than the collisions between particles of different sizes.
We will discuss the sticking rate quantitatively in Section \ref{sec4:application}.

\subsection{Comparison with the V\"olk-type model}

\begin{figure}[t]
\begin{center}
\includegraphics[scale=.6]{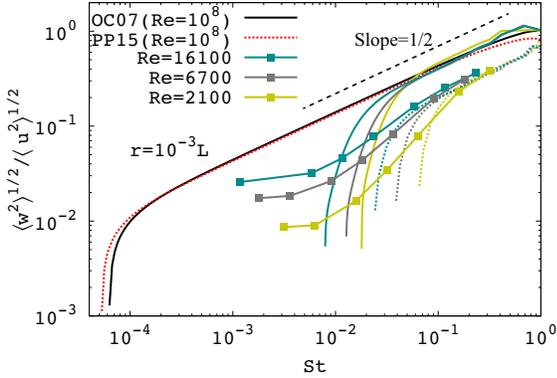}
\end{center}
\caption{rms relative velocity at $r=10^{-3}L$ is compared to the V\"olk-type model. 
Colored dashed lines with points represent the DNS results at $t=3T$
for $Re=2100$ (yellow), $Re=6700$ (gray), and $Re=16100$ (cyan). 
Colored lines denote the V\"olk-type model employing $E(k)$ in the DNSs. 
Black line shows the V\"olk-type model by OC07.
Red dotted line shows an improved closed expression by PP15 (their Eq. 25). 
Colored dotted lines show an improved formula of V\"olk-type models presented in PP15 (at their Fig.10).
\label{fig:wrms3}}
\end{figure}

As for the relative velocities of particles induced by turbulence, a closed-form expression developed by 
OC07 has been widely adopted \citep[e.g., a review by][]{Johansen2014}.
The original formalism (the V\"olk-type model) was developed by \cite{Volk1980} and \cite{Markiewicz1991}.
\cite{Cuzzi2003} obtained closed-form expressions for the V\"olk-type model.
OC07 generalized the approach and results of \cite{Cuzzi2003} to obtain closed-form expressions
for relative velocities between particles of arbitrary, and unequal, sizes.

For equal-size particles with a stopping time of $\tau_{\rm p}$, the closed-form expression 
(OC07) is given by
\begin{eqnarray}
 \langle w^2 \rangle=\int^\infty_{k^*}4E(k)(1-F(k)^2)dk, \label{omega}
\end{eqnarray}
where $F(k)=\tau_{\rm p}/(\tau_{\rm p}+\tau_k(k))$, $\tau_k(k) = [2E(k)k^3]^{-1/2}$, and 
the critical wavenumber $k^*$ for the particle with stopping time $\tau_{\rm p}$ is determined by
\begin{equation}
{1\over \tau_{\rm p}}={1\over \tau_k(k^*)}+k^*v_{\rm rel}(k^*),
\label{critical}
\end{equation}
where the relative velocity $v_{\rm rel}$ between a particle and an eddy is given by
\begin{equation}
v_{\rm rel}^2(k^*)=\int_{k_L}^{k^*}2E(k')[\tau_{\rm p}/(\tau_{\rm p}+\tau_k(k'))]^2dk'
\end{equation}
and $k_L$ is the smallest wavenumber.
OC07 assumed the energy spectrum given by 
\begin{equation}
E(k) = \left\{
\begin{array}{ll}
 V_{\rm g}^2/(3k_L)(k/k_L)^{-5/3}&\hbox{for}\ k_L<k<k_\eta, \\
 0\ &\hbox{for}\ k<k_L, \ k>k_\eta,
\end{array}
\right.
\label{inertialrange}
\end{equation}
for which the total energy is $V_{\rm g}^2/2$.
As shown in PP15, in turbulent flows with a wide inertial range, the above model predicts a $\tau_{\rm p}^{1/2}$ scaling for $\langle w^2\rangle^{1/2}$, if $\tau_{\rm p}$ is in the inertial range.

In Figure \ref{fig:wrms3}, we compare the DNS results of the relative velocity with the V\"olk-type model given by Equation (\ref{omega}) in which Equation (\ref{inertialrange}) is used for $E(k)$ by setting $k_\eta/k_L=Re^{4/3}$ and $Re=10^8$. For comparison, the V\"olk-type models which tentatively employ the DNS data for $E(k)$ in Equation (\ref{omega}) are also plotted. 
In the range $St=O(0.1)$, the latter models agree well with the V\"olk-type model that assumes the model spectrum (\ref{inertialrange}).
We can recognize that the DNS results for $Re=16100$ has a slope $1/2$ for $0.05\lesssim St\lesssim 0.2$. The slope is consistent with the V\"olk-type model for high-$Re$ turbulence. However, the DNS values are smaller than the V\"olk-type model by a factor of two. PP15 did not show the clear slope of the relative velocity, but their result suggested that the V\"olk-type models typically overestimate the rms of the particle relative velocity.
Our results suggest that in the inertial range of high-$Re$ turbulence the rms of the particle relative velocity obeys the scaling law, but the values are smaller than the V\"olk-type model by a factor of two. 
PP15 proposed an improved closed expression of the relative velocity (their Eq. 25) and  an improved formula of V\"olk-type models (at their Fig.10).
We plot these expressions in Figure  \ref{fig:wrms3}  for comparison. The difference between the former and the V\"olk-type model by OC07 is small.
On the other hand, the latter formula seem to work well at $St\gtrsim 0.1$.

\begin{figure*}[t]
\begin{center}
\includegraphics[scale=1]{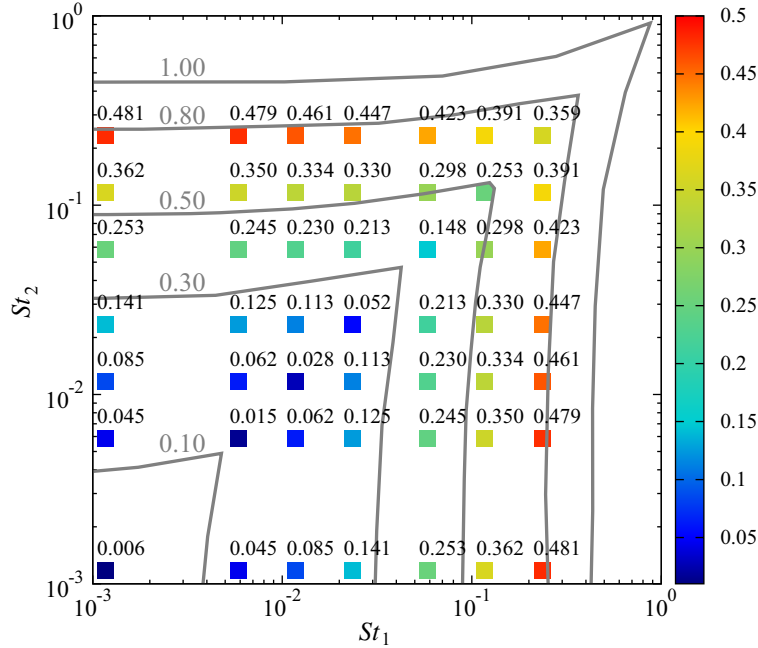}
\end{center}
\caption{Turbulence-induced rms relative velocities $\langle w^2 \rangle^{1/2}$ normalized to $\langle u^2\rangle ^{1/2}$
between two particles characterized by $St_1$ and $St_2$. Colored squares (with rms values) represent the DNS results 
for a particle separation of $r=\eta/4$ at $t=3T$ ($Re=16100$). 
Gray contours (with attached rms values) denote the prediction of the V\"olk-type model (\ref{ppvel}), 
which is equivalent to Figure 4C of OC07.
The diagonal line $(St_1=St_2)$ corresponds to Figure \ref{fig:wrms3}. 
Note that the DNS results for $St_1$ (or $St_2$) $<0.05$ are more or less affected by viscosity.
\label{Figppvel}}
\end{figure*}

In Figure \ref{Figppvel}, we compare the DNS results of the relative velocities to those of the closed-form expression 
derived by OC07. The DNS results for $Re=16100$ are shown by colored squares with rms values 
as functions of $St_1=\tau_1/t_L$ and $St_2=\tau_2/t_L$. 
For unequal-sized particles with stopping times of $\tau_1$ and $\tau_2$, the closed-form expression (OC07) is given by
\begin{equation}
\begin{split}
\langle w^2 \rangle=&{V_{\rm g}^2\over t_L}(\Bigl[\tau_k+{\tau_1^2\over \tau_1+\tau_k}\Bigr]_{\tau_k=t_\eta}^{\tau_k={\rm max}(\tau_1^*,\tau_2^*)} \\
& +{\tau_2-\tau_1\over \tau_1+\tau_2}\Bigl[{\tau_1^2\over \tau_1+\tau_k}\Bigr]_{\tau_k={\rm max}(\tau_1^*,\tau_2^*)}^{\tau_k=t_L} \\
& +(1\leftrightarrow 2)),
\label{ppvel}
\end{split}
\end{equation}
where $t_L=(V_Lk_L)^{-1}$, $V_L^2=(2/3)V_{\rm g}^2$, and $t_\eta=Re^{-1/2}t_L$. $\tau_1^*$ and $\tau_2^*$ are the solutions of Equation (\ref{critical}) for $\tau_{\rm p}=\tau_1$ and $\tau_{\rm p}=\tau_2$, respectively. 
The relative velocity variance $\langle w^2 \rangle ^{1/2}$ by this formula is shown by contours 
in Figure \ref{Figppvel} for the case $Re=10^8$.
The comparison with the relative velocities obtained by the DNS 
clarifies that the DNS results are smaller than half of the values of the closed-form expression given by OC07, 
regardless of the $St$ ratios. 
In particular, for equal-sized particles with $St \lesssim 10^{-2}$,
the DNS results are smaller by an order of magnitude. These reduced relative velocities have a significant impact on
avoiding the fragmentation barrier, which is discussed in details in Section \ref{sec4:application}.

\subsection{Collision kernel}
\label{collision_kernel}

\begin{figure}[b]
\begin{center}
\includegraphics[scale=.6]{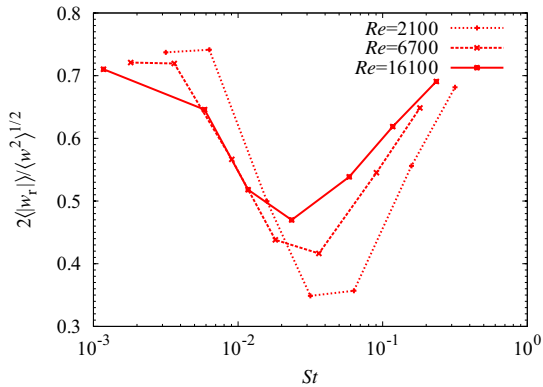}
\caption{Averaged radial relative velocities $\langle|w_{\rm r}|\rangle$ at $r=10^{-3}L$, normalized by the relative velocity variance,
as a function of the $St$ number. 
Results are shown for $Re=16100$ (solid line), $Re=6700$ (dashed line), and $Re=2100$ (dotted line). 
\label{fig:wrandw1}}
\end{center}
\end{figure}

So far, we have focused on the variance of relative velocities,
which gives a measure of the collision rate. Here, we consider
the collision kernel to incorporate the turbulent clustering effect properly. 
The collision rate per unit volume between two particles with radii $a_1$ and $a_2$ can be expressed as 
\begin{equation}
n_1(a_1)n_2(a_2)\Gamma(a_1, a_2),
\end{equation}
where $n_1(a_1)$ and $n_2(a_2)$ are the average number densities and $\Gamma(a_1, a_2)$ is the collision kernel.
The kernel formula hitherto used in dust coagulation models is $\Gamma^{\rm com}=\pi d^2\langle w^2\rangle^{1/2}$
with $d=a_1+a_2$, where the effect of turbulent clustering is not taken into account
and the rms relative velocity, $\langle w^2\rangle^{1/2}$, is usually taken from the model of \cite{Volk1980}.
As a statistical mechanical description of $\Gamma$ for zero-inertia particles, \cite{Saffman1995} proposed 
a spherical formulation, in which
the kernel is given by $\Gamma=2\pi d^2\left<|w_{\rm r}|\right>$ with 
the radial relative velocity, $w_{\rm r}={\bm w}\cdot({\bm X}_2-{\bm X}_1)/|{\bm X}_2-{\bm X}_1|$.
In addition, \cite{Sundaram1997} 
considered the turbulent clustering effect on $\Gamma$ and derived an expression for finite-inertia particles. 
\cite{Wang2000} suggested that a formula (based on the spherical formulation),
\begin{equation}
\Gamma^{\rm sph}=2\pi d^2\langle|w_{\rm r}|\rangle g(d), \label{gamma_sph} 
\end{equation}
is more accurate than the formula considered by \cite{Sundaram1997}, where
$g(d)$ is the radial distribution function (RDF) at $r=d$.
The RDF is related to the two-point correlation function $\xi(r)$ by $g(r)=1+\xi(r)$,
and evaluated by
\begin{equation}
g(r)={N(r)\over 4\pi r^2 \bar{n}dr},
\end{equation}
where $N(r)$ is the average number of particles in a spherical shell of volume $4\pi r^2 dr$ 
at a distance $r$ from a reference particle and $\bar{n}$ is the average particle number density.
For a uniform distribution of the particles, we have $g(r)=1$, i.e., $\xi(r)=0$.
For a non-uniform distribution of the particles due to turbulent clustering, we have $g(r)>1$.
Recently, \cite{Pan2014_IV} and PP15 evaluated $\Gamma^{\rm sph}$
by performing the DNS with $Re=O(10^3)$ to study the effect of turbulent clustering on the collision kernel. 

Here, we use our DNS data with the Reynolds numbers up to $Re=16100$ to
obtain $\langle|w_{\rm r}|\rangle$ and $g(r)$, and elucidate the $Re$-dependence of the collision kernel.
Figure \ref{fig:wrandw1} shows the normalized radial relative velocity, $2\langle|w_{\rm r}|\rangle/\langle w^2\rangle^{1/2}$,
as a function of $St$ for different $Re$ numbers.
As seen in this figure, the values are always less than unity, 
and therefore the use of $\langle w^2\rangle^{1/2}$ for the collision kernel leads to the overestimation.
Additionally, since the values are a decreasing or increasing function of $St$, 
$\langle|w_{\rm r}|\rangle$ does not obey the power law $(\propto St^{1/2})$ 
like $\langle w^2\rangle^{1/2}$ does (Figure \ref{relative_velocities2}).
Therefore, it is concluded that there are quantitative 
and qualitative discrepancies from $\langle w^2\rangle^{1/2}$ 
for evaluating the collision kernel $\Gamma$.
The low value of $\langle|w_{\rm r}|\rangle$ leads to the enhancement of the particle sticking rates
as discussed in Section \ref{sec4:application}.
\cite{Pan2014_IV} studied the $St$-dependence of the ratio $2\langle|w_{\rm r}|\rangle/\langle w^2\rangle^{1/2}$ 
and showed that the values of the ratio approach its Gaussian value of $(8/3\pi)^{1/2}\simeq 0.92$ for $St>1$. 
They argued that the non-Gaussianity peaks at $St_\eta\simeq 1$, resulting in a dip in the ratio at $St_\eta\simeq 1$. 
Figure \ref{fig:wrandw1} shows that 
the bottom position of the dip is around $St_\eta\simeq1-2$, and therefore
the dip becomes shallower accordingly as $Re$ increases.

\begin{figure}[t]
\begin{center}
\includegraphics[scale=.65]{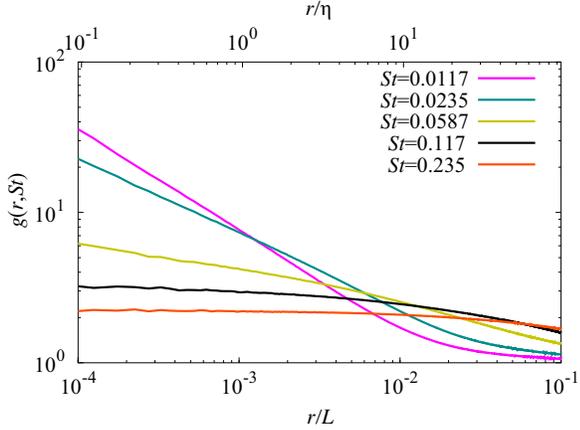}
\caption{RDFs for particles with different $St$ numbers; each value of $St$ corresponds 
respectively to $St_\eta=1, 2, 5, 10,$ and $20$. The used data are $512^3$ particles at $t = 3T$ in the DNS of Run2048-1 ($Re = 16100$).
\label{fig:rdf1}}
\end{center}
\end{figure}

\begin{figure}[t]
\begin{center}
\includegraphics[scale=.6]{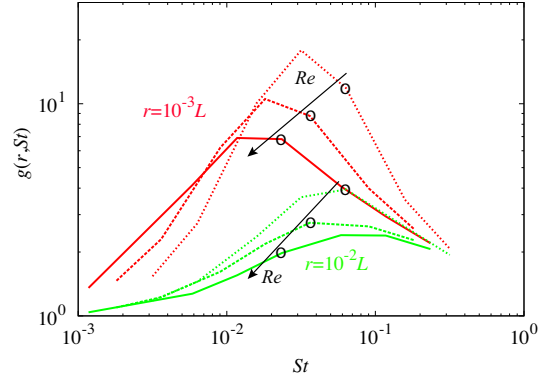}
\caption{$St$-dependence of the RDFs 
at $r =10^{-3}L$ (red) and $2\times 10^{-3}L$ (green) for different $Re$ numbers: 
$Re=16100$ (solid line), $Re=6700$ (dashed line), and $Re=2100$ (dotted line). 
The circles denote the data for $St_\eta=2.0$.
\label{fig:rdf2}}
\end{center}
\end{figure}

\begin{figure}[t]
\begin{center}
\includegraphics[scale=.6]{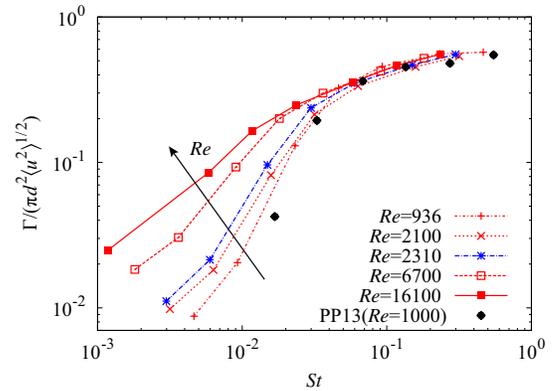}
\end{center}
\caption{Collision kernel per unit cross section in the spherical formulation at distance $d=10^{-3}L$ for different $Re$ numbers.
The values are measured at $t=3T$ and normalized by $\langle u^2 \rangle ^{1/2}$ for each run. 
The data from PP13 ($Re=1000$, $r=\eta/4\sim 2\times 10^{-3}L$) are also plotted
for comparison.
\label{fig:collisionkernel}}
\end{figure}

Figure \ref{fig:rdf1} shows the $St$-dependence of $g(r)$ in the DNS of Run 2048-1 ($Re = 16100$).
The behaviors of $g(r)$ are quantitatively consistent with the recent DNS results by \cite{Ireland2016a}.
In Figure \ref{fig:rdf1}, we see that $g(r)$ is approximately constant at $r/L \lesssim 10^{-2}$ 
for $St \gtrsim 0.1$. This fact allows us to use the constant value when considering the collision statistics
(as discussed in Section \ref{sec4:application}). 

Figure \ref{fig:rdf2} shows the $St$-dependence of $g(r)$ at $r/L=10^{-3}$ and $r/L=10^{-2}$ for three different values of $Re$. 
Here, $r/L=10^{-3}$ $(10^{-2})$ corresponds to $r/\eta=1/4$ $(5/2)$, $1/2$ $(5),$ and $1$ $(10)$ for $Re=2100, 6700,$ and $16100$, 
respectively. 
Figure \ref{fig:rdf2} demonstrates that $g(r)$ has a peak for a fixed value of $r$. 
Also, Figure \ref{fig:rdf1} suggests that the value of $St$ at the peak depends on $r$,
and $g(r)$ has a peak near $St_\eta=1$ for small values of $r=O(\eta)$.
In Figure \ref{fig:rdf2}, we notice that, for fixed values of $St=O(0.1)$ and $r/L$, $g(r)$ is a decreasing function of $Re$ 
provided that $St_\eta\geq 2.0$; this $Re$-dependence is weaker for the larger values of $St$.
We observe also that, for fixed values of $St=O(0.1)$ and $Re$, $g(r)$ becomes larger as $r/L$ decreases; 
this $r/L$-dependence is weaker for the larger values of $St$ (as shown in Figure \ref{fig:rdf1}).
The $Re$-dependence and $r/L$-dependence of $g(r)$ suggest that 
an asymptotic behavior of $g(r)$ (in the range of $St=O(0.1)$) for much smaller $r$ 
and for much higher $Re$ can be surmised from the DNS data of $Re=O(10^4)$.

\cite{Ireland2016a} showed that, for fixed values of $St_\eta\gtrsim 1$ and $r/\eta$, $g(r)$ is an increasing function of $Re$. 
Their $Re$-dependence does not contradict with the results in Figure \ref{fig:rdf2}. 
In fact, our DNS data also gives almost the same $Re$-dependence as that shown in Figure 21 of \cite{Ireland2016a}. 
However, it should be noted that their collision statistics are completely different from each other.

Figure \ref{fig:collisionkernel} presents the $Re$-dependence of the resulting collision kernel obtained by our DNS.
In Figure \ref{fig:collisionkernel}, we see a noticeable increase in the collision kernel in the small $St$ range 
with increasing $Re$. For example, the value of the collision kernel 
at $St=2\times 10^{-2}$ for $Re=16100$ is six times larger than that for $Re=1000$.
However, the Reynolds number dependence of the collision kernel at $St\approx 0.02-0.2$ 
seems to converge provided that $Re\gtrsim 10^4$. 
From this finding, it is expected that the collision kernel at $St\approx 0.02-0.2$ 
for a much higher Reynolds number ($Re \gg 10^4$)  is similar to  
those obtained by our DNS of $Re=O(10^4)$.

\begin{figure}[t]
\begin{center}
\includegraphics[scale=.6]{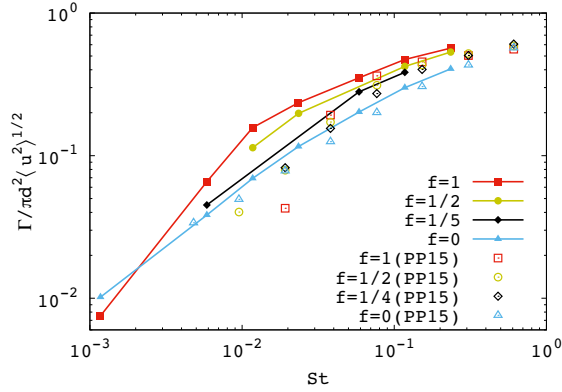}
\caption{Collision kernel per unit cross section in the spherical formulation at the distance of $\eta/4$ 
for particle pairs with fixed Stokes ratios, $f\equiv St_2/St_1=1$, 
$1/2$, $1/5$, and $0$. The data are obtained at $t=3T$ in the DNS of $Re=16100$. 
The data from PP15 ($Re = 1000$, $r = \eta/4\sim2\times10^{-3}L$) are also plotted for comparison.
\label{fig:bid2}}
 \end{center}
\end{figure}

As for bidisperse cases, 
the $St(=St_1)$-dependence of the collision kernel (Equation (\ref{gamma_sph})) is presented
for different $f= St_2/St_1$ in Figure \ref{fig:bid2}. 
To compare the collision kernel at the scale of the smallest eddies, we show the results
at the distance of $\eta/4$.
As seen in this figure, 
the values of $\Gamma$ for different-sized particles are smaller than those for the identical particles.
This result is caused by the fact that the concentration of particles 
due to turbulent clustering occurs more effectively for the identical particles than for the different-sized particles.
The comparison with the data from PP15 shows that the DNS results for $St\gtrsim 0.1$ at $Re=16100$ are not far from those at $Re=1000$.

\subsection{PDF of radial relative velocity}
\label{vrPDF}

In the collision kernel, we have adopted the averaged value of the radial relative 
velocities, $\langle|w_{\rm r}|\rangle$. However, to assess the fraction of particles which have
velocities lower than the critical collision velocity, we should derive 
the PDF of $w_{\rm r}$. 
Since the critical collision velocity may be different between equal-sized collisions and
different-sized ones, we need to obtain the PDF depending on the Stokes number ratio $f\equiv St_2/St_1$.

\begin{figure}[t]
\begin{center}
\includegraphics[scale=.65]{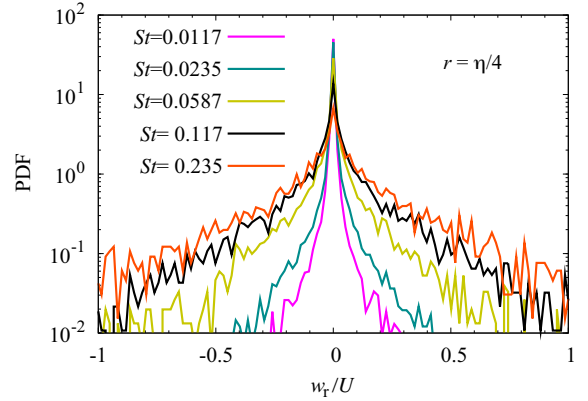}
\caption{PDF of the normalized radial component of the relative velocity for each $St$.
The relative velocities are measured at $t=3T$ for the pairs of equal-sized particles at a separation of $r=\eta/4$ 
in the DNS of $Re=16100$.\label{fig:pdf1}
}
\end{center}
\end{figure}

Based on the DNS data of $Re=16100$, we acquire the PDF for equal-sized particles, $P^{({\rm eq})}$.
Figure \ref{fig:pdf1} shows the resultant PDF of the normalized radial relative velocity $w_{\rm r}/U$ 
at a separation of $r=\eta/4$ for each $St$ number, where
$U$ is the rms value of the fluctuating velocity in one direction. 
The negative and positive values of $w_{\rm r}/U$ represent the approaching and receding pairs, respectively. 
The variance and kurtosis of $x\equiv w_{\rm r}/U$ are defined by 
\begin{equation}
V\equiv\langle (x-\langle x\rangle)^2 \rangle
\label{varian}
\end{equation}
and
\begin{equation}
K\equiv\langle (x-\langle x\rangle)^4 \rangle/\langle (x-\langle x\rangle)^2\rangle^2,
\label{kurtosis}
\end{equation}
respectively, and listed in Table \ref{fitting_parameters}.
It is shown that when $St\geq 1.17\times 10^{-2}(St_\eta\geq 1.0)$, the variance is an increasing function of $St$ and the kurtosis is a decreasing function of $St$.
The data in Table \ref{fitting_parameters} indicate that 
the values of $V$ and $K$ are respectively approximated by 
\begin{equation}
V\approx St^{1.3}\ \hbox{and}\ K\approx 3.0+0.79\times St^{-1.3},
\label{approxVK}
\end{equation}
in the range $5.87\times 10^{-2}\leq St\leq 2.35\times 10^{-1}$.

\begin{figure}[t]
\begin{center}
\includegraphics[scale=.65]{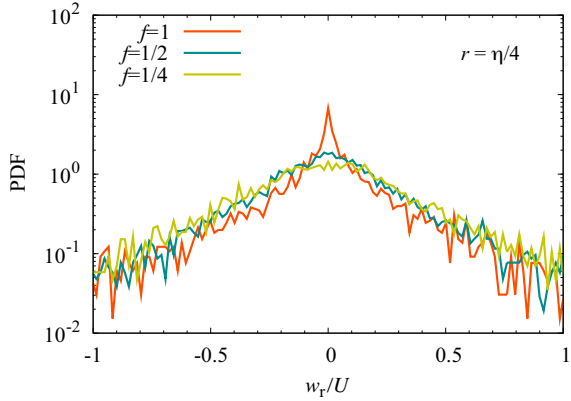}
\caption{Same as Figure \ref{fig:pdf1} but for the pairs of the different-sized particles $St_1$ and $St_2$,
where $St_1=0.235$ and $f=St_2/St_1$.
\label{fig:bidpdf}
}
\end{center}
\end{figure}

The PDFs for different-sized particles, $P^{({\rm diff})}$, are shown for $f\equiv St_2/St_1=1$, $1/2$, and $1/4$, fixing $St_1=0.235$,
in Figure \ref{fig:bidpdf}. 
As expected, Figure \ref{fig:bidpdf} suggests that the non-Gaussianity of the PDFs is weakened as $f$ decreases 
(as the size difference becomes large). 
The similar trend has already been observed in low-$Re$ simulations (see, e.g., Figs. 2, 4, and 8 in \cite{Pan2014_III}).
Our DNS results confirm that the trend is also true for higher $Re$ turbulence.
The variance and the kurtosis of the PDFs for the different-sized particles ($f\neq 1$) 
are listed in Table \ref{fitting_parameters_bidisparse}.
The data in Tables \ref{fitting_parameters} and \ref{fitting_parameters_bidisparse} imply that, as $f$ decreases, the value of $K$ monotonically decreases 
while the variance monotonically increases. 

\begin{table*}[t]
\caption{%
Variance ($V$) and kurtosis ($K$) of the normalized relative velocity ($w_{\rm r}/U$) measured at $t=3T$ for the equal-sized particles with a separation $r=\eta/4$ in the DNS of $Re=16100$.
Parameters, $\mu$ and $\beta$, in Equation (\ref{Px}) determined by $K_{\rm se}=K$ and $V_{\rm se}=V$ are also listed.
}
\begin{center}
\begin{tabular}{cccccc} \hline\hline
$St(St_\eta)$        & 0.0117(1.0) & 0.0235(2.0) & 0.0587(5.0) & 0.117(10.0) & 0.235(20.0) \\ 
\hline
Variance($V$) & 8.04E-4 &2.58E-3 & 2.21E-2 & 6.15E-2  & 1.39E-1      \\
Kurtosis($K$) & 283    & 85.4    & 34.1    & 13.8   & 8.33          \\
$\mu$        & 0.273   & 0.352   & 0.453   & 0.632  & 0.813     \\
$\beta$  & 1.72E-5  & 3.17E-4  & 4.75E-3 & 3.32E-2 & 1.01E-1       \\
 \hline
 \label{fitting_parameters}
\end{tabular}
\end{center}
\end{table*}

\begin{table}[t]
\caption{Same as Table \ref{fitting_parameters} but for the pairs of different-sized particles. The value of $f$ denotes the ratio $St_2/St_1$.}
\begin{center}
\begin{tabular}{cccc} \hline\hline
$f$ & $1/2$ & $1/4$ & $1/2$ \\ 
$St_1(St_{\eta 1})$ & 0.117(10.0) & 0.235(20.0) & 0.235(20.0) \\ 
$St_2(St_{\eta 2})$ & 0.0587(5.0) & 0.0587(5.0)  & 0.117(10.0) \\ \hline
Variance($V$) & 8.91E-2 & 1.74E-1 & 1.50E-1  \\
Kurtosis($K$) & 9.40  & 5.13    & 7.03     \\
$\mu$ & 0.760  & 1.12  & 0.899  \\
 $\beta$ & 0.069 & 0.203 & 0.130  \\
 \hline
 \label{fitting_parameters_bidisparse}
\end{tabular}
\end{center}
\end{table}

Following \cite{Sundaram1997,Wang2000}; PP13; \cite{Pan2014_III},
we attempt to fit the PDF of $x=w_r/U$ with a stretched exponential function given by
\begin{equation}
P_{\rm se}(x)={\mu\over 2\beta\Gamma(1/\mu)}\exp\left[-\left({|x|\over \beta}\right)^\mu\right],
\label{Px}
\end{equation}
where $P_{\rm se}(x)$ is normalized to satisfy $\int_{-\infty}^\infty P_{\rm se}(x) dx =1$ as in PP13 and \cite{Pan2014_III}.
For this function, the variance and kurtosis are analytically given by 
\begin{equation}
V_{\rm se}=\beta^2\Gamma(3/\mu)/\Gamma(1/\mu)
\end{equation}
and
\begin{equation}
K_{\rm se}={\Gamma(1/\mu)\Gamma(5/\mu)\over\Gamma(3/\mu)^2},
\end{equation}
respectively. Since the value of $K_{\rm se}$ does not depend on the value of $\beta$, 
we can first determine the value of $\mu$ with $K_{\rm se}=K$ and then the value of $\beta$ with $V_{\rm se}=V$, 
where $V$ and $K$ are the DNS values. Specifically, the formula to determine the values 
of $\mu$ and $\beta$ is given by
\begin{equation}
K_{\rm se}=K, \quad
\beta=\left[{\Gamma(1/\mu)V\over \Gamma(3/\mu)}\right]^{1/2},
\label{parameters_for_model}
\end{equation}
where $V$ and $K$ are given by the approximated formulas (Equation (\ref{approxVK})).
The values of $\mu$ and $\beta$ determined by the DNS for the case $Re=16100$ 
are shown in Tables \ref{fitting_parameters} and \ref{fitting_parameters_bidisparse}.

\begin{figure}[t]
\begin{center}
\includegraphics[scale=.65]{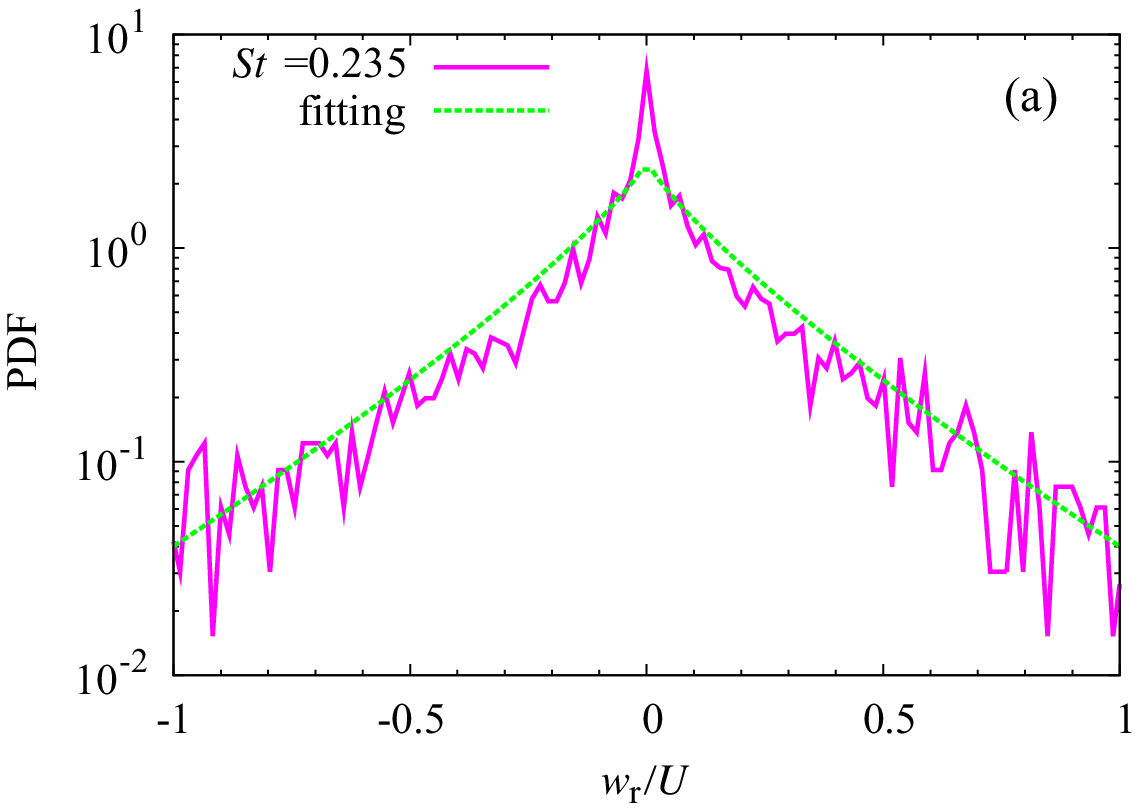}
\includegraphics[scale=.65]{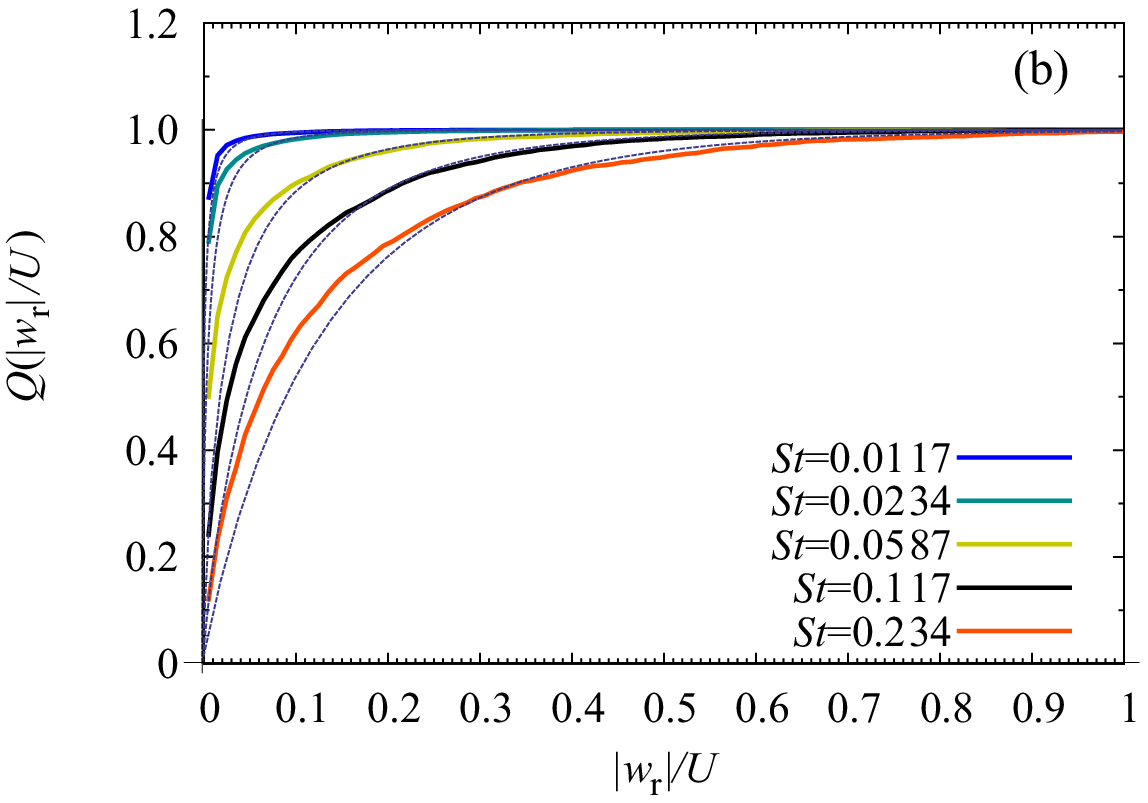}
\caption{(a) Comparison between the PDF of the normalized radial relative velocities for $St=0.235$ 
at a separation of $r=\eta/4$ and the fitting function (Equation (\ref{Px})). 
(b) Values of the integral $Q(x)\equiv\int_{-x}^x P^{({\rm eq})}(x) dx$ (solid lines) 
for different $St$, where $x=|w_{\rm r}|/U$ and $P^{({\rm eq})}(x)$ is the PDF from the DNS of $Re=16100$.
The data are compared with $\int_{-x}^x P^{({\rm eq})}_{\rm se}(x) dx$ (dotted lines), 
where $P^{({\rm eq})}_{\rm se}(x)$ is given by Equation (\ref{Px}), in which the values of $\beta$ and $\mu$ in Table \ref{fitting_parameters} are used.}
\label{fig:fitpdf}
\end{center}
\end{figure}

Theoretically, a stretched exponential PDF with $\mu=4/3$ was predicted for inertial-range particles under the assumption of exactly Gaussian flow velocity and Kolmogorov scaling (Gustavsson et al. 2008; PP13).
The values of $\mu$ for the inertial-range particles of $St=O(0.1)$ in our DNS are less than the theoretical value and suggest that non-Gaussian behavior of the flow velocity affects on the PDF of the radial relative velocity of the inertial range particles.
Hence, the fitting function (Equation (\ref{Px})) and the values of $\mu$ and $\beta$ in Tables \ref{fitting_parameters} and \ref{fitting_parameters_bidisparse} are useful for modeling the PDF of the radial relative velocities for the inertial range particles.
Note that the values of $\mu$ for the particles of $0.1<St<0.3$ in our DNS are not far from those for the corresponding $St$ values reported for low-$Re$ simulations in \cite{Pan2014_III}. This fact suggests that not only the variance but also the PDF of the normalized relative velocity for $St\gtrsim 0.1$ is not so sensitive to the values of the Reynolds number.

An example of the stretched exponential fit to the PDF is plotted in Figure \ref{fig:fitpdf}(a) for the case of identical particles. 
We confirm that the stretched exponential function qualitatively approximates well the PDFs of the relative velocities.
In Figure \ref{fig:fitpdf}(b), we show the accuracy of the fitting quantitatively by evaluating the integral $Q(x)\equiv\int_{-x}^x P^{({\rm eq})}(x) dx$.

\section{Implications for Planetesimal Formation}
\label{sec4:application}

The density of the compact dust aggregates is virtually equal to that of the monomer 
($\rho_{\rm s}\approx 1$g cm$^{-3}$).
The formation of cm-sized ($St=0.01-0.1$) compact aggregates (pebbles) is a key for
the growth to planetesimals via the streaming instability \citep[e.g.][]{Johansen2014}.
On the other hand, dust aggregates can have fluffy structures with much lower bulk densities 
if the compression by impacts is sufficiently weak 
\citep{Wada2009,Okuzumi2009,Okuzumi2012,Zsom2010, Zsom2011}.
In recent years, the collisions of fluffy dust aggregates have been explored 
by performing $N$-body molecular-dynamics simulations 
\citep{Wada2009,Paszun2009,Wada2013,Meru2013,Seizinger2013,Gunkelmann2016}.
These studies have revealed that fluffy aggregates, depending on the breaking energy, 
may resolve the difficulties due to the bouncing and fragmentation barriers.
So far, fluffy aggregates are thought to likely form as a result of the coalescence of icy grains.
However, quite recently, \cite{Arakawa2016} have shown that 
fluffy aggregates can result from the collisions of nanometer-sized silicate grains. 
Here, according to \cite{Johansen2014}, we consider two cases of
compact aggregates with $\rho_{\rm s}=1$g cm$^{-3}$ and extremely fluffy aggregates with $\rho_{\rm s}=10^{-5}$g cm$^{-3}$.

As for a protoplanetary disk, we assume an $\alpha$-model, in which the turbulent viscosity is given as $\nu_{\rm t}=\alpha c_{\rm s}H$, 
$c_{\rm s}$ is the sound speed, $H$ is the vertical scale height, and the typical value of $\alpha$ is between $\sim 10^{-4}$ to $10^{-2}$. 
In addition, we employ the Minimum-Mass Solar Nebula (MMSN) model (Hayashi 1981), 
which provides the gas temperature
$T_{\rm g} = 280{(R/\rm{AU})^{ - 1/2}}{\rm{K}}$, 
the gas density 
$\rho_{\rm g}=1.2 \times 10^{-9}$ $(R/{\rm AU})^{-11/4}$g cm$^{-3}$,
the sound speed
$c_{\rm s}=1.1\times 10^3(R/{\rm AU})^{-1/4} {\rm m}\ {\rm s}^{-1}$, 
the surface mass density of gas
$\Sigma_{\rm g} = 1.7 \times 10^3 (R/\rm{AU})^{-3/2} \rm{g\ cm}^{-2}$,
and the vertical scale height of the gas
$H=4.7\times 10^{-2} (R/{\rm AU})^{5/4} {\rm AU}$,
where $R$ is the distance from the central star.
Assuming a cross section of $2.5\times 10^{-15}{\rm cm}^2$ for hydrogen molecules, we have the kinematic viscosity given by $\nu=6.0\times 10^4(R/{\rm AU})^{5/2}{\rm cm}^2\ {\rm s}^{-1}$.
In the $\alpha$-model, the characteristic velocity and the characteristic length scale in the turbulence are given by $U=\alpha^{1/2}c_{\rm s}$ and $L=\alpha^{1/2}H$, respectively, i.e., 
$U=1.1\times10^2(\alpha/10^{-2})^{1/2}(R/{\rm AU})^{-1/4}\ {\rm m}\ {\rm s}^{-1},$
$L=5.5\times10^{8}(\alpha/10^{-2})^{1/2}(R/{\rm AU})^{5/4}\ {\rm m}.$
Thus, the Reynolds number $Re$ is given by
$Re=UL/\nu\sim10^{10}(\alpha/10^{-2})(R/{\rm AU})^{-3/2}$
and the Kolmogorov length and time scales are estimated using $\eta=Re^{-3/4}L$ and $\tau_\eta=Re^{-1/2}(L/U)$.
Turbulence characteristics for $\alpha=10^{-2}$ and $10^{-4}$ are listed in Table \ref{tab4}.

\begin{table}[h]
\begin{center}
\caption{Turbulence characteristics in the $\alpha$-models of the protoplanetary disk at $R=1$AU.}
\begin{tabular}{cccccc} \hline\hline
$\alpha$		   & $U$[m s$^{-1}$]     & $L$[m]        & $Re$ & $\eta$[m] & $\tau_\eta$[s] \\ \hline
 $10^{-2}$& $110$    	 & $5.5 \times 10^{8}$ & $10^{10}  $   & $17$ & 50\\
 $10^{-4}$& $11$    	 & $5.5 \times 10^{7}$ & $10^{8}  $   & $54$ & 500\\ \hline
\label{tab4}
\end{tabular}
\end{center}
\end{table}

\begin{figure*}[t]
\begin{center}
\includegraphics[scale=.55]{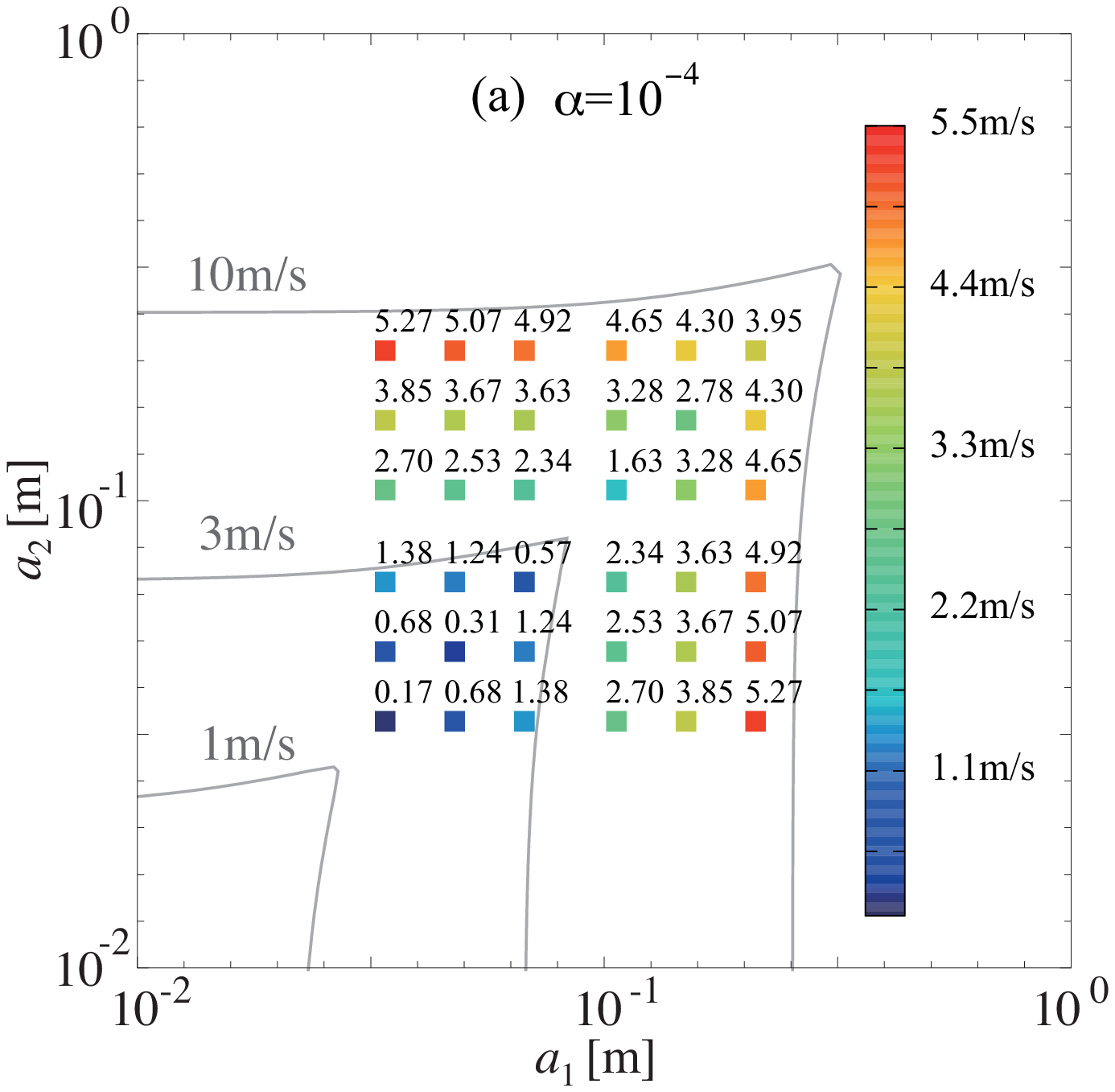}\quad\quad\ 
\includegraphics[scale=.55]{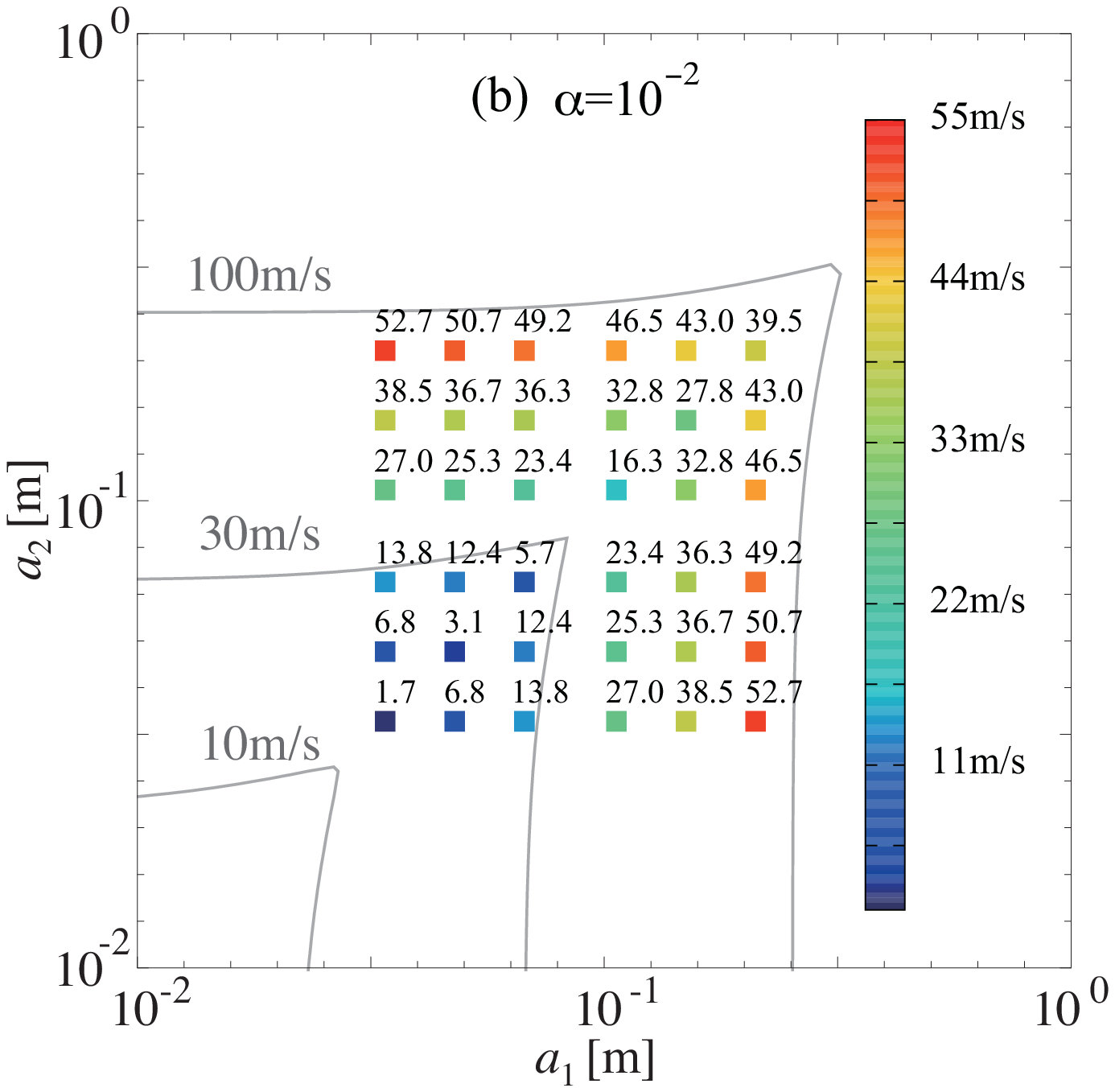}\\
\vskip0.5cm
\includegraphics[scale=.54]{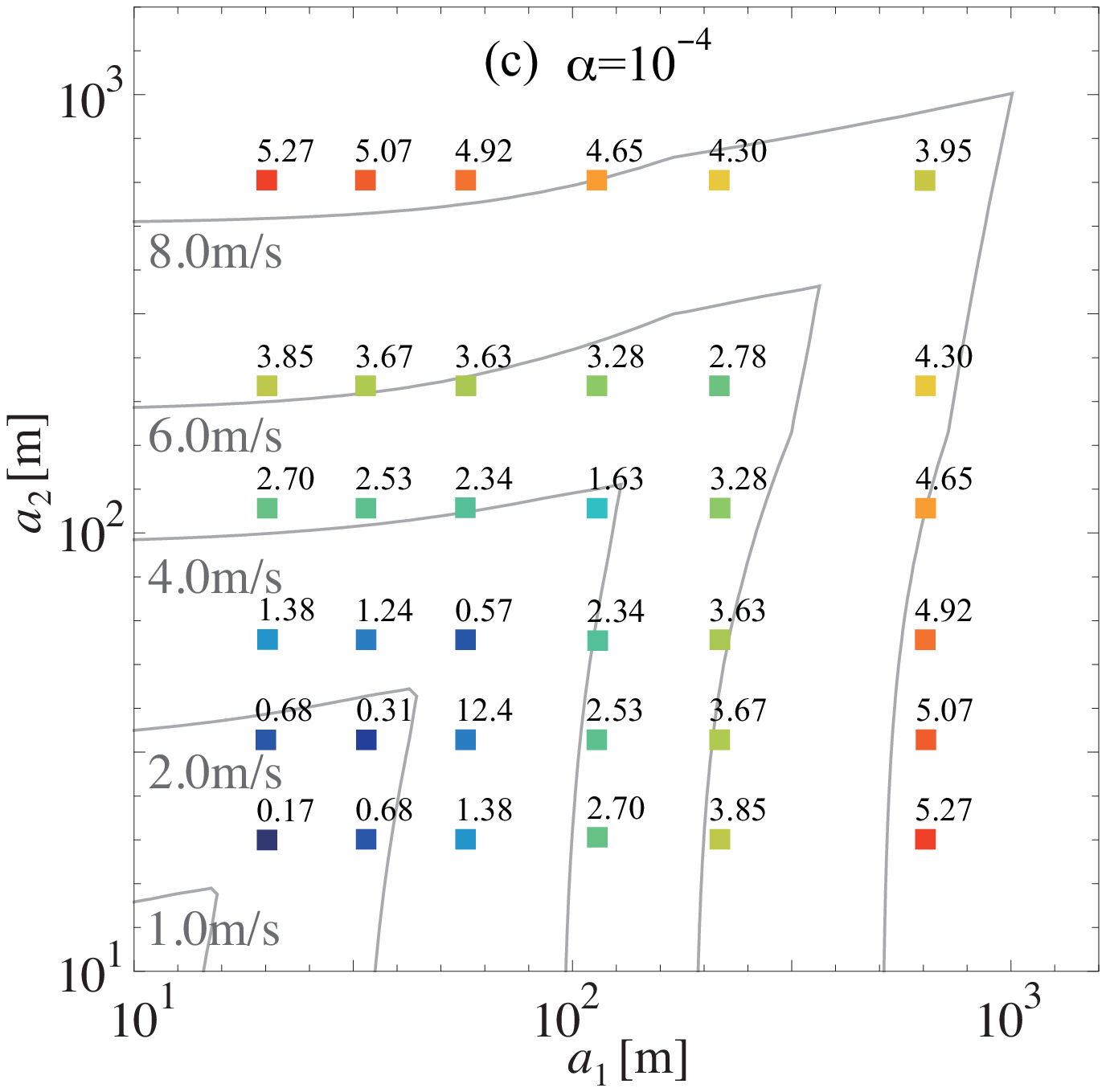}\qquad\quad
\includegraphics[scale=.54]{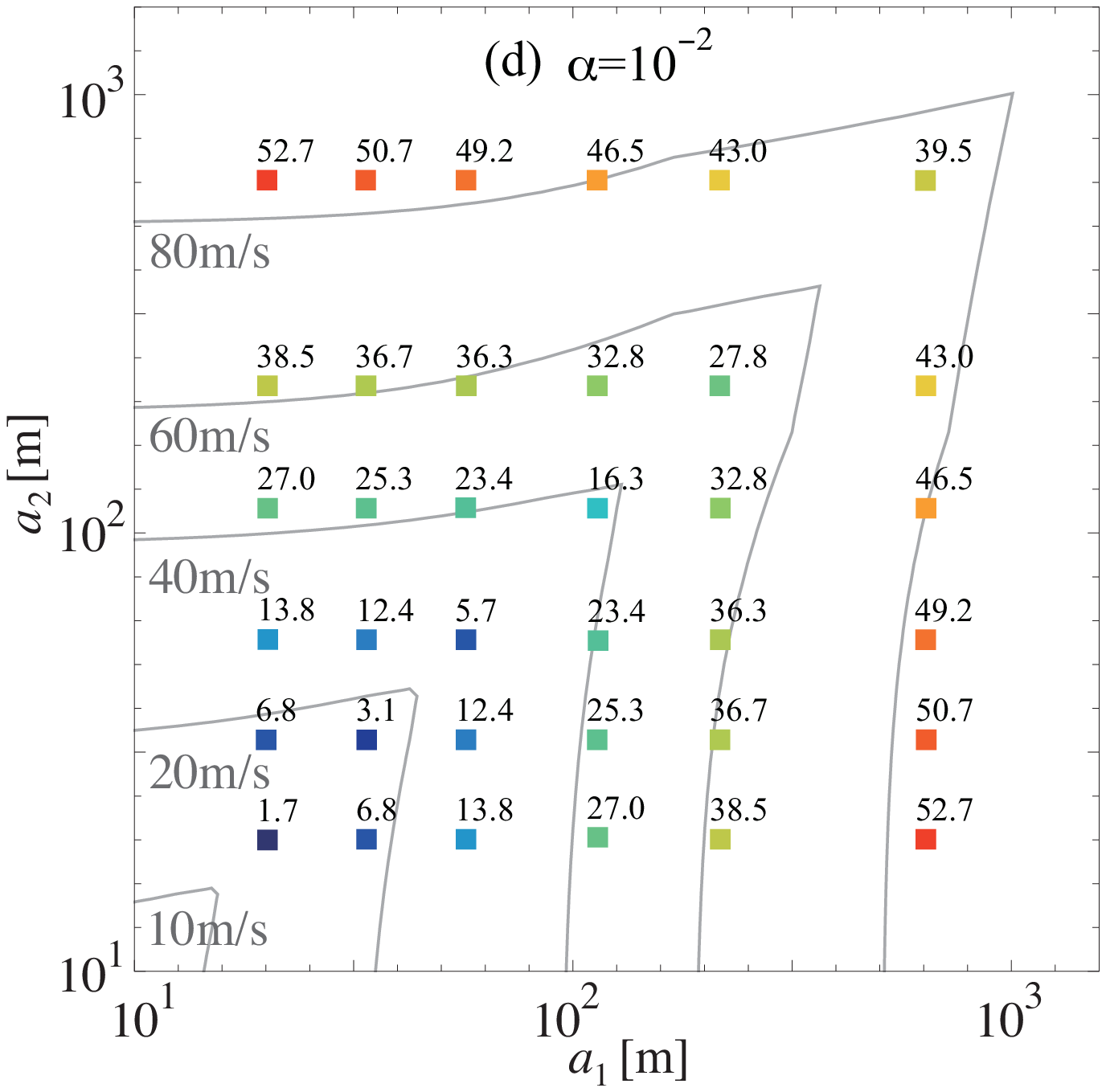}
\end{center}
\caption{rms relative velocities (in m s$^{-1}$) between two particles of different sizes for 
compact aggregates ($\rho_{\rm s}=1{\rm g}\ {\rm cm}^{-3}$) (top panels)
and fluffy aggregates ($\rho_{\rm s}= 10^{-5}{\rm g}\ {\rm cm}^{-3}$) (bottom panels).
Here, the MMSN model is employed assuming 1 AU from the central star for (a, c) $\alpha=10^{-4}$ and (b, d) $\alpha=10^{-2}$.
Filled squares (with rms values) represent the DNS results, while the
contours (with rms values) are the theoretical estimates obtained from OC07, 
using the particle sizes corresponding to the Stokes numbers given by \cite{Johansen2014}.
}
\label{Jophansen}
\end{figure*}

\subsection{Collision velocity in the MMSN model}
Based on the MMSN model, we evaluate the rms relative velocities induced by turbulence. 
In Figure \ref{Figppvel}, we showed the rms relative velocities as a function of $St$ number pairs. 
Translating the Stokes numbers into the particle sizes according to \cite{Johansen2014},
we can obtain the rms collision velocities for compact and fluffy dust aggregates.
In Figure \ref{Jophansen}, we show the resultant rms collision velocities at 1 AU from the central star,
for $\alpha=10^{-4}$ in panels (a, c) and for $\alpha=10^{-2}$ in panels (b, d). 
The upper panels are the results for compact aggregates, and the lower are those for fluffy aggregates.
For comparison, the estimates derived with the closed-form analytic formula by OC07 
are also depicted. 
We find that the rms relative velocities by the DNSs are smaller by more than a factor of two,
compared to the prediction of the closed-form expression. 
In particular, for particles of equal size, the DNS results are much lower. 
These results have a considerable impact on the sticking rates and the collision timescale as discussed below.

\begin{figure*}[t]
\begin{center}
%
\includegraphics[scale=1.]{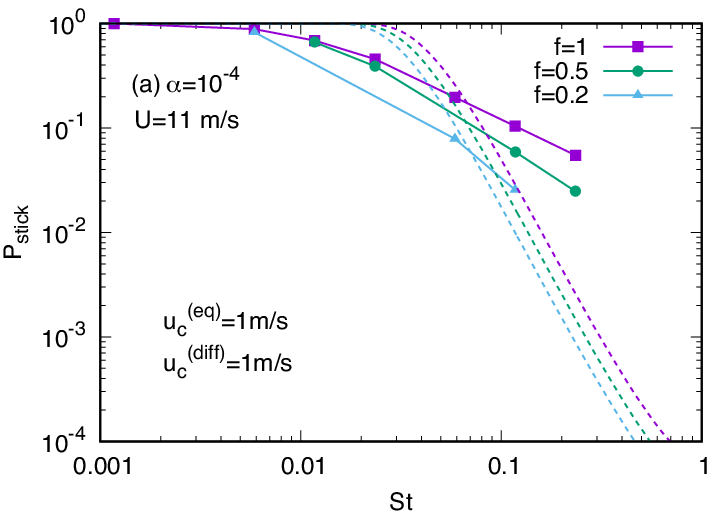}\quad\quad
\includegraphics[scale=1.]{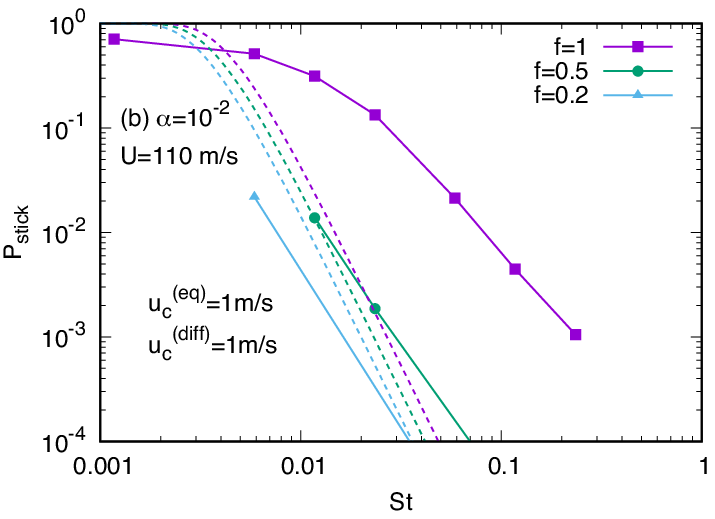}
\includegraphics[scale=1.]{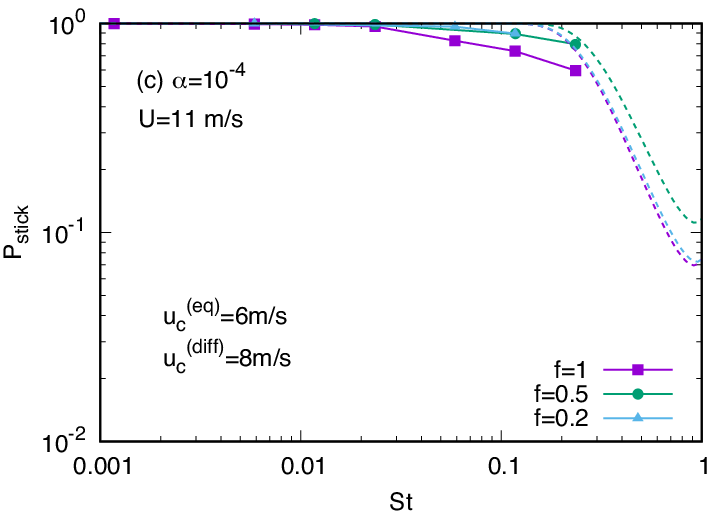}\quad\quad
\includegraphics[scale=1.]{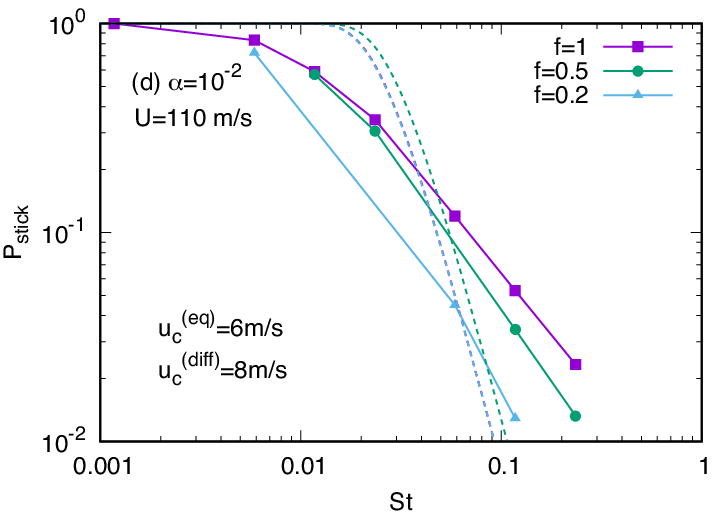}
\end{center}
\caption{Sticking rates of colliding particles as a function of $St$, depending on $f=St_2/St_1$.
The MMSN model is employed, assuming 1 AU from the central star for (a, c) $\alpha=10^{-4}$ ($U=11$m s$^{-1}$) 
and (b, d) $\alpha=10^{-2}$ ($U=110$m s$^{-1}$).
In the top panels (a, b), the critical collision velocity is assumed to be
$u_{\rm c}^{({\rm eq})}=u_{\rm c}^{({\rm diff})}=1$m s$^{-1}$, while in the bottom panels (c, d) 
those are $u_{\rm c}^{({\rm eq})}=6$m s$^{-1}$ for equal-sized collisions and $u_{\rm c}^{({\rm diff})}=8$m s$^{-1}$ for different-sized collisions, which are given for silicate dust by \cite{Wada2013}. 
Colored solid curves represent the DNS results calculated from the PDFs, 
using pairs of particles at a separation $r=\eta/4$ in the DNS of $Re=16100$. 
Colored dotted curves denote the theoretical prediction assuming a Gaussian (Maxwell) distribution 
whose variance is given by OC07. 
}
\label{fig:Sticking_rates}
\end{figure*}

\vskip 0.5cm
\subsection{Sticking rates of colliding pairs} 
We estimate the ``sticking rates", which is defined to be the probabilities 
of the sticking of colliding pairs per unit time. 
Hubbard (2012) pointed out that to estimate the sticking rates of colliding pairs 
the weighting factor proportional to the collision rate should be taken into account.
In the cylindrical kernel formulation, the collision rate is proportional to $|{\bm w}|$, 
and the collision-rate weighted distribution of the 3D amplitude is obtained
 from the unweighted PDF, $P(|{\bm w}|)$, simply by
\begin{equation}
 P_{cyl}(|{\bm w}|)=|{\bm w}|P(|{\bm w}|)/\left<|{\bm w}|\right>. \label{weightedPDF}
\end{equation}
As shown in \cite{Pan2014_IV},  the spherical kernel formulation of the weighted PDF is also possible but is more complicated. 
Therefore, we use (\ref{weightedPDF}) for estimating the sticking rates, which are given by
\begin{equation}
P_{\rm stick}^{({\rm eq, diff})}= \int_0^{u_{\rm c}^{({\rm eq,diff})}} P^{({\rm eq,diff})}_{cyl}(x) dx
\end{equation}
for equal-sized and different-sized collisions, respectively, where $x=|{\bm w}|$ and 
$u_{\rm c}$ is the critical velocity above which 
the collisions lead to the bouncing or fragmentation (see e.g. \cite{Wada2013}).

\cite{Wada2009}, using $N$-body molecular dynamics simulations, have obtained
the critical velocity for silicate aggregates to be
$u_{\rm c}=1.1 (a/0.76 \mu{\rm m})^{-5/6}$m s$^{-1}$, where $a$ is the size of monomers in aggregates.
It agrees well with the laboratory experiments \citep{Blum2000}.
Additionally, \cite{Wada2013} have derived a scaling relation of
the critical collision velocity as a function of the breaking energy.
They have considered fluffy aggregates composed of ballistic particle-cluster aggregation clusters,
which are fairly compact (fractal dimension $D \sim 3$). 
Even in such compact aggregates, all surface interactions between monomers in contact in the aggregates
determine the critical collision velocity, which depends on the size of the monomers. 
For icy aggregates, they have derived the critical collision velocity as
\begin{eqnarray}
u_{\rm c}^{({\rm eq})}&=&60 (a/\mu{\rm m})^{-5/6} {\rm m\ s}^{-1}, \\
u_{\rm c}^{({\rm diff})}&=&80 (a/\mu{\rm m})^{-5/6} {\rm m\ s}^{-1}, 
\end{eqnarray}
for collisions between equal-sized and different-sized particles, respectively, 
and for silicate aggregates
\begin{eqnarray}
u_{\rm c}^{({\rm eq})}&=&6 (a/\mu{\rm m})^{-5/6} {\rm m\ s}^{-1}, \label{wada_eq_silicate}\\ 
u_{\rm c}^{({\rm diff})}&=&8 (a/\mu{\rm m})^{-5/6} {\rm m\ s}^{-1}. \label{wada_dif_silicate}
\end{eqnarray}
Recently, \cite{Gunkelmann2016} studied the porosity-dependence of the 
fragmentation of fluffy aggregates composed of silicate grains of 0.76 $\mu$m radius, and found that
the critical velocity for agglomerate fragmentation decreases with the porosity of the aggregates.
Although the critical collision velocities are still under debate, those for silicate aggregates
are smaller by an order of magnitude than those for icy aggregates. 
Therefore, the sticking rates are expected to be much lower in silicate aggregates. 
Here, we consider two cases of $u_{\rm c}$ for silicate aggregates;
$u_{\rm c}=1$m s$^{-1}$ for compact aggregates
and equations (\ref{wada_eq_silicate}) and (\ref{wada_dif_silicate})
for fluffy aggregates.

We evaluate the sticking rates of $P_{\rm stick}^{({\rm eq})}$ and $P_{\rm stick}^{({\rm diff})}$ assuming $R=1$ AU
for the cases with $\alpha=10^{-4}$ and $10^{-2}$, based on the PDFs obtained by the DNS with $Re=16100$. 
In Figure \ref{fig:Sticking_rates}, we plot the results as a function of $St$.
For comparison, we also plot the theoretical prediction 
based on the collision-rate weighted distribution of the 3D amplitude
assuming a Gaussian (Maxwell) distribution with the variance given by OC07.

Panels (a) and (b) show respectively the DNS results for $\alpha=10^{-4}$ and $10^{-2}$, assuming
$u_{\rm c}=1$m s$^{-1}$. 
Quite interestingly, the sticking rates are not strongly dependent on $\alpha$ if $St\lesssim 0.01$
and keep a high level of $\gtrsim 50\%$, although the theoretical prediction from a Gaussian distribution
declines steeply for $\alpha=10^{-2}$.  
Besides, the declination of the rates at $St \gtrsim 0.01$ is
much more gradual compared to the theoretical prediction, especially in the case of $\alpha=10^{-2}$.
It shows that the non-Gaussianity of the velocity distribution function
due to turbulent clustering makes 
the sticking rates for $St\gtrsim 0.01$ remarkably higher than those theoretically expected.
Also, it is worth noting that the rates for equal-sized particles ($f=1$) are 
higher than those for different-sized particles ($f \ne1$).
In the case of $\alpha=10^{-2}$, the difference is more than an order. 
This comes from the fact that the variance of the relative velocities is smaller 
for equal-sized particles ($f=1$), as shown in Figure \ref{Jophansen}. 
These results imply that equal-sized particles grow much faster than different-sized particles,
and therefore the fraction of equal-sized particles increases with time. 

Panels (c) and (d) show respectively the DNS results for $\alpha=10^{-4}$ and $10^{-2}$, assuming
the critical collision velocity for fluffy aggregates given by \cite{Wada2013}. 
In panel (c), the DNS results show that the sticking rates are approximately unity at $St\lesssim 0.03$, 
and also the sticking rates are slightly higher for different-sized particles ($f \ne1$).
As shown in Figure \ref{fig:bidpdf}, the PDFs have small differences for different $f$ values. 
However, $u_{\rm c}^{({\rm diff})}$ is larger than $u_{\rm c}^{({\rm eq})}$, and therefore 
$P_{\rm stick}^{({\rm diff})}$ is a bit higher than $P_{\rm stick}^{({\rm eq})}$. 
But, in the case of $\alpha=10^{-2}$, the rates for $f=1$ are higher than those for $f \ne1$
and decline more gradually compared to the theoretical prediction at $St\gtrsim 0.01$.

\begin{figure*}[t]
\begin{center}
%
%
%
\includegraphics[scale=1]{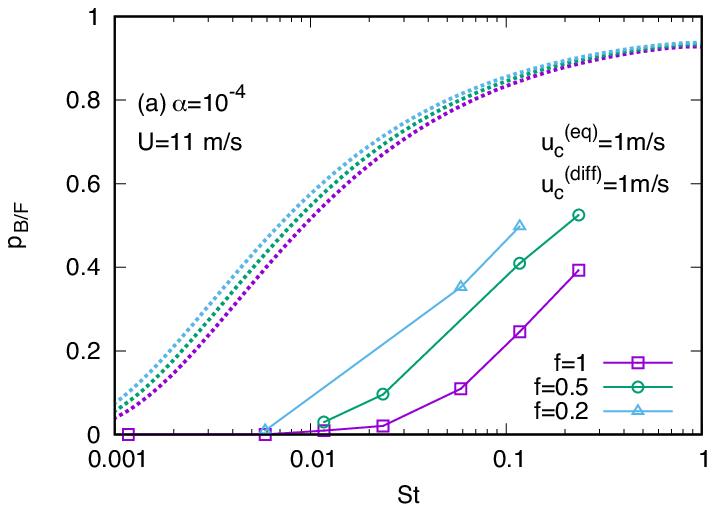}\quad\quad
\includegraphics[scale=1]{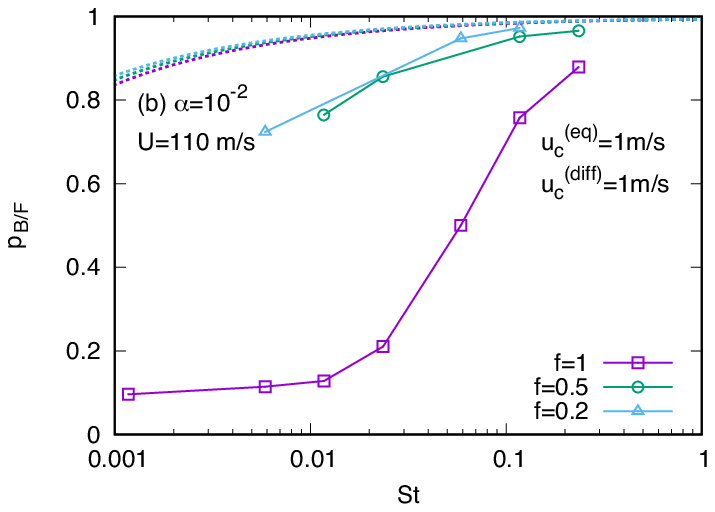}
\includegraphics[scale=1]{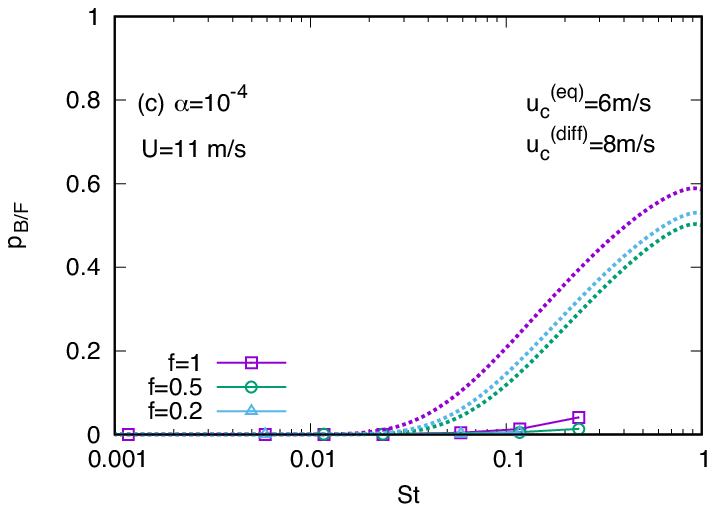}\quad\quad
\includegraphics[scale=1]{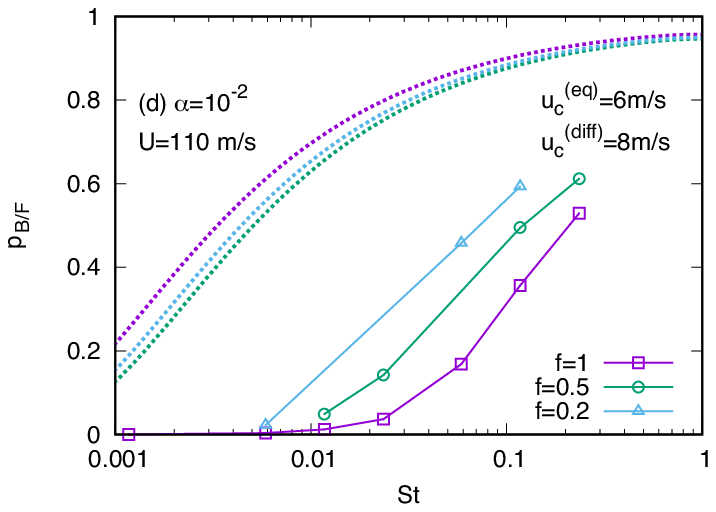}
\end{center}
\caption{ Bouncing/fragmentation probabilities of colliding particles
as a function of $St$, depending on $f=St_2/St_1$.  
The meanings of curves are the same as those in Fig. \ref{fig:Sticking_rates}.
}
\label{fig:bouncing}
\end{figure*}

\subsection{Bouncing/fragmentation probabilities}  
Using the PDFs, we can evaluate the bouncing/fragmentation fraction of colliding particles, 
if we specify the critical collision velocity $u_{\rm c}$ as in the previous subsection.
The bouncing/fragmentation probabilities are estimated as 
\begin{equation}
p_{\rm B/F}^{({\rm eq,diff})}=\int_{-\infty}^{-u_{\rm c}^{({\rm eq,diff})}/U} P^{({\rm eq,diff})}(x) dx
\end{equation}
for equal-sized and different-sized collisions, respectively, where $x=w_r/U$.

We evaluate the B/F probabilities of $p_{\rm B/F}^{({\rm eq})}$ and $p_{\rm B/F}^{({\rm diff})}$ 
assuming $R=1$ AU for the cases with $\alpha=10^{-4}$ and $10^{-2}$, based on the PDFs obtained by the DNS with $Re=16100$. 
In Figure \ref{fig:bouncing}, we plot the results as a function of $St$.
In any case, the obtained B/F probabilities are significantly lower than 
the theoretical prediction from a Gaussian distribution.
Also, importantly, the B/F probabilities for equal-sized particles ($f=1$) are 
much lower than those for different-sized particles ($f \ne1$). 
For $St\lesssim 0.01$, the B/F probabilities for $f=1$ are lower than 10\%
even in the case of $\alpha=10^{-2}$.
As discussed in the above, the rms relative velocity in the DNSs is smaller than
that in the closed-form expression, and also the averaged radial relative velocity 
is even lower than the rms relative velocity, as shown in Section \ref{collision_kernel}.
Owing to such low values of relative velocities and the non-Gaussianity of the PDF 
in the DNSs, the B/F probabilities are not dramatically enhanced even for $\alpha=10^{-2}$. 
Hence, equal-sized particles preferentially survive even in a highly turbulent flow.

\begin{table*}[t]
\begin{center}
\caption{Collision timescale of dust particles defined by $t_{\rm coll}\equiv 1/[n_{\rm d} g(r) \langle |w_{\rm r}|\rangle \sigma]$, 
where $n_{\rm d}(=\rho_{\rm d}/m_{\rm d})$ is the number density of the dust particles with the radius $a$ 
($\rho_{\rm d}$ is the dust density in the disk and $m_{\rm d}$ the mass of the dust aggregate), $g(r)$ is the RDF of the particle, 
$\langle |w_{\rm r}|\rangle$ is the average of the radial relative velocities, 
and $\sigma(=\pi (2a)^2)$ is the cross-section area. 
The timescale is shown
for compact aggregates ($\rho_{\rm s}=1$g cm$^{-3}$) and fluffy aggregates ($\rho_{\rm s}=10^{-5}$g cm$^{-3}$).
\label{tab5}
}
\begin{tabular}{c|ccccc|ccccc} 
\multicolumn{1}{c}{ }&\multicolumn{5}{c}{Compact ($\rho_{\rm s}=1$g cm$^{-3}$)} & \multicolumn{5}{c}{Fluffy ($\rho_{\rm s}=10^{-5}$g cm$^{-3}$)} \\ \hline
$St$ & $0.0117$ & $0.0235$&$0.0587$&$0.117$&$0.235$&$0.0117$&$0.0235$&$0.0587$ & $0.117$ & $0.235$ \\ \hline
 $a$[cm] &$4.36$&$6.17$& $9.75$   & $13.8$   & $19.5$ & $1379$&$1950$&$3083$   & $4361$   & $6167$ \\
 $m_{\rm d}$[kg] & $0.347$&$0.982$&$3.88$ & $11.0$ & $31.1$ &$110$&$311$&$1228$ & $3473$ & $9823$ \\ 
$t_{\rm coll}$[yr] & $1.07$&$1.01$&$1.07$&$1.12$&$1.32$ & $0.00338$&$0.00319$&$0.00337$&$0.0035$&$0.00418$ \\ \hline
\end{tabular}
\end{center}
\end{table*}

\begin{table}[t]
\begin{center}
\caption{Maximal density contrast of dust particles as a function of the Stokes number ($St$).
$\Delta L$ is a coarse-grain scale in units of $\alpha^{1/2}H$, 
where the density is calculated in the volume of $\Delta L^3$.}
\label{tab6}
\begin{tabular}{c|ccccc} \hline
\multicolumn{1}{c|}{$\Delta L$} & \multicolumn{5}{c}{$St$} \\ 
$[ \alpha^{1/2}H ]$  & 0.0117 & 0.0234 & 0.0587 & 0.117 & 0.234 \\ \hline
0.16  & 1.89 & 2.46 & 3.92 & 5.12 & 6.01 \\ 
0.32  & 1.45 & 1.76 & 2.63 & 3.49 & 4.00 \\ 
0.64  & 1.23 & 1.36 & 1.56 & 1.88 & 2.30 \\ \hline
\end{tabular}
\end{center}
\end{table}

\subsection{Streaming instability}

As shown above, the fluffy silicate aggregates may coalesce with
a high probability in a weakly turbulent disk ($\alpha =10^{-4}$). 
However, the coagulation of compact aggregates are much less effective 
for $St \gtrsim 0.1$ in a highly turbulent disk ($\alpha =10^{-2}$) , 
and hence there is still an obstacle of the radial drift barrier. 
For the growth from cm-sized ($St=0.01-0.1$) compact aggregates (pebbles) 
to planetesimals, 
the streaming instability may be a possible route to circumvent this obstacle.

\citet{YG05} discovered the streaming instability promoted by the action-reaction pair of the
drag force between solid particles and gas, and the growth of pebbles via 
the streaming instability has been extensively explored \citep[][and references therein]{Johansen2014}.
Importantly, \citet{Johansen2009} and \citet{BS2010}  have pointed out that
there is a critical solid abundance $Z_c$, above which spontaneous strong concentration of solids occurs,
where $Z$ is the solid-to-gas surface mass density ratio, $Z\sim 0.01$ being the solar abundance. 
Using 2D simulations, \citet{Carrera2015} found that
the critical abundance is super-solar and increases drastically with decreasing
particle size for $St<0.1$. 
Very recently, using 2D and 3D high-resolution simulations, \citet{Yang2017} have shown that
the critical abundance is not a steep function of $St$, and 
$0.01 <Z_c < 0.02$ for particles of $St= 10^{-2}$ and 
$0.03 < Z_c < 0.04$  for particles of $St= 10^{-3}$. 
Although these are slightly super-solar, some mechanisms are still required to
enhance the solid abundance to allow the formation of planetesimals via the streaming instability.

In our simulations, as shown in Figure \ref{fig:rdf1}, 
the solid abundance is enhanced by the turbulent clustering, dependent on $St$. 
Since the clustering is dependent on scales, the density contrast of dust particles,
$\rho_d/\bar{\rho}_d$, depends on a coarse-grain scale.
In Table \ref{tab6}, we show the maximal density contrast of dust particles
as a function of $St$, 
where $\Delta L$ is a coarse-grain scale in units of $\alpha^{1/2}H$ 
and the density is calculated in the volume of $\Delta L^3$.
As seen in Table \ref{tab6}, the enhancement of solid abundance becomes larger 
according as $St$ increases and $\Delta L$ lessens. 
If $\alpha =10^{-2}$ and $\Delta L \le 0.03H$, the enhancement is a factor of $\sim 1.5$
for  $St =0.01$ and $\sim 3.5$ for  $St =0.1$.
This satisfies the condition for the critical abundance 
shown in Fig. 9 of \citet{YG05}.

As shown in Figure \ref{fig:Sticking_rates},
the sticking of equal-sized particles is faster than that of different-sized ones.
This supports the assumption of particles of the same size employed by \citet{Yang2017}.
The sticking rate of equal-sized particles is as high as $\gtrsim 50\%$ at $St \lesssim 0.01$, and 
reduces to $\lesssim 10\%$ at $St \gtrsim 0.1$, in the case of $\alpha =10^{-2}$. 
Furthermore, Figure \ref{fig:bouncing} shows that the bouncing/fragmentation probabilities are
as low as $\lesssim 10\%$ at $St \lesssim 0.01$, and increase steeply
toward $St \gtrsim 0.1$. 
Therefore, equal-sized particles of $St \sim 0.01$ are expected to selectively grow. 

Also, we evaluate the collision time-scale of dust particles, which is given by 
$t_{\rm coll}\equiv 1/[n_{\rm d} g(r) \langle |w_{\rm r}|\rangle \sigma]$, 
where $n_{\rm d}$ is the number density of the dust particle of size $a$, $g(r)$ is the RDF of the particle, 
$\langle |w_{\rm r}|\rangle$ is the averaged radial relative velocity, 
and $\sigma(=\pi (2a)^2)$ is the cross-section area. 
Note that the collision timescale is independent of the value of $\alpha$,
since $n_{\rm d} \propto \alpha^{-1/2}$ and $\langle |w_{\rm r}|\rangle \propto \alpha^{1/2}$.
As for $g(r)$ and $\langle |w_{\rm r}|\rangle$, we use the values at $r/\eta=1/4$ for $Re=16100$. 
If we specify the values of $St$ and $\rho_{\rm s}$, we have $n_{\rm d}$ and $a$
in the Stokes regime, using the vertical scale-height for dust particles \citep{Okuzumi2012}. 
In Table \ref{tab5}, the evaluated collision times are listed 
for compact aggregates ($\rho_{\rm s}=1$g cm$^{-3}$) and fluffy aggregates ($\rho_{\rm s}=10^{-5}$g cm$^{-3}$).
Since $g(r)$ is a decreasing function of $St$ and $\langle |w_{\rm r}|\rangle$ is an increasing function, 
the collision timescale is insensitive to $St$ and much shorter than the drift timescale
even for compact aggregates. 

Considering the above assessments, we can expect that 
mostly equal-sized particles of $0.01 \lesssim St \lesssim 0.1$
grow in a timescale of $\sim \Omega ^{-1}$. 
Simultaneously, the turbulent clustering enhances the solid abundance.
According to \citet{Yang2017}, the critical solid abundance tends to be minimal toward $St \sim 0.1$.
Hence, the streaming instability may be triggered at $0.01<St<0.1$, and
the strong concentration of solids proceeds in a timescale of $\gtrsim 100 \Omega ^{-1}$.

\section{Conclusions}
In order to investigate the dynamical statistics of dust particles 
through turbulent clustering in a protoplanetary disk,
we have performed the high-resolution DNSs of the Navier-Stokes equations. 
The number of grid points and the Reynolds number 
are up to $2048^3$ and $Re=16100$, respectively, which are of the highest resolution ever
in astrophysical DNSs. These large-scale DNSs 
have allowed us to track the motion of dust particles with Stokes numbers of 
$0.01\lesssim St \lesssim 0.2$
in the inertial range for the first time. 
As results of these simulations, we have found the following:

\begin{itemize}

\item As the Reynolds number of the turbulence increases (or the inertial range extends),
the rms relative velocity, $\langle w^2\rangle^{1/2}$, of particle pairs at a fixed small separation (normalized by the integral length scale) is augmented for small $St$ number particles, and is asymptotically proportional to $St^{1/2}$  in the inertial range.

\item The rms relative velocities by the DNSs are smaller by more than a factor of two, 
compared to those from the closed-form expression derived by \cite{Ormel2007}, 
irrespective of the $St$ number ratios of the particle pairs. 
Also, the averaged radial relative velocity is even lower than the rms relative velocity. 
Hence, the findings by
\citet{Pan2013,Pan2015} have been confirmed by high-$Re$ DNSs. 

\item The PDFs of the radial relative velocities are highly
non-Gaussian, and are well fitted by a stretched exponential function like Equation (\ref{Px}).
The PDF of the normalized relative velocity for $St\gtrsim0.1$ is not so sensitive to the values of the Reynolds number.
Hence, the results are consistent with those at low-$Re$ by \citet{Pan2014_III}.

\item
Almost independently of $\alpha$, the sticking rates of colliding particles 
are as high as $\gtrsim 50\%$ for particles
of $St\lesssim 0.01$ and declines gradually at $St \gtrsim 0.01$, 
although the theoretical prediction from a Gaussian distribution
declines steeply for $\alpha=10^{-2}$.  
This comes from the non-Gaussianity of the radial relative velocities and the smaller variance of the 
relative velocities as a result of the turbulent clustering. 

\item
Since the variance of the relative velocities for equal-sized particles ($f=1$) 
is smaller than that for different-sized particles ($f \ne1$), 
the sticking rates for $f=1$ are higher than those for $f \ne1$. The difference is
larger than an order of magnitude in the case of $\alpha=10^{-2}$. 
It implies that equal-sized particles grow much faster than different-sized particles,
and therefore the fraction of equal-sized particles increases with time. 

\item
The bouncing/fragmentation probabilities are significantly lower than 
the theoretical prediction from a Gaussian distribution.
The probabilities for equal-sized particles ($f=1$) are 
much lower than those for different-sized particles ($f \ne1$),
in the case of $\alpha=10^{-2}$. 
Hence, equal-sized particles preferentially survive 
even in a highly turbulent flow. 

\item 
The turbulent clustering enhances the solid abundance.
The enhancement in a scale of 0.03 scale-height of the disk 
increases from a factor of $\sim 1.5$ at $St =0.01$ to $\sim 3.5$ at  $St =0.1$, 
in the case of $\alpha=10^{-2}$. 
Therefore, the streaming instability may be triggered at $0.01<St<0.1$.

\end{itemize}

In the present DNSs, we have assessed sticking rates, but the actual coagulation 
of dust particles has not been incorporated. Hence, we cannot predict the mass function of dust aggregates
as a result of the hierarchical coagulation. This is a significant issue for planetesimal formation. 
In DNSs in the near future, we will explore the growth of dust aggregates in turbulence.

\acknowledgments

We are grateful to K. Furuya, E. Kokubo, S. Michikoshi, T. Nakamoto, S. Okuzumi, and K. Yoshida for their valuable discussions.
The computational resources of the K computer provided by the RIKEN Advanced 
Institute for Computational Science through the HPCI System Research project (Projects ID: hp150174, ID: hp160102, and ID: hp170087) was partially used in this study. 
It also used the FX100 system at the Information Technology Center, Nagoya University. 
This research was supported in part by the Interdisciplinary Computational Science Program of the Center for Computational Sciences, University of Tsukuba, Grant-in-Aid for Scientific Research (B) by JSPS (15H03603,15H03638), and MEXT as "Exploratory Challenge on Post-K computer" (Elucidation of the Birth of Exoplanets [Second Earth] and the Environmental Variations of Planets in the Solar System).


\begin{thebibliography}{}
\bibitem[Adachi et al.(1976)]{Adachi1976} Adachi, I., Hayashi, C., \& Nakazawa, K.\ 1976, Progress of Theoretical Physics, 56, 1756 
\bibitem[Arakawa \& Nakamoto(2016)]{Arakawa2016} Arakawa, S., \& Nakamoto, T.\ 2016, \apjl, 832, L19 
\bibitem[Bai \& Stone(2010)]{BS2010} Bai, X.-N., \& Stone, J.~M.\ 2010, \apj, 722, 1437 
\bibitem[Birnstiel et al.(2009)]{Birnstiel2009} Birnstiel, T., Dullemond, C.~P., \& Brauer, F.\ 2009, \aap, 503, L5 
\bibitem[Birnstiel et al.(2012)]{Birnstiel2012} Birnstiel, T., Klahr, H., \& Ercolano, B.\ 2012, \aap, 539, A148 
\bibitem[Blum \& Wurm(2000)]{Blum2000} Blum, J., \& Wurm, G.\ 2000, \icarus, 143, 138
\bibitem[Blum \& Wurm(2008)]{Blum2008} Blum, J., \& Wurm, G.\ 2008, \araa, 46, 21 
\bibitem[Bragg \& Collins(2014)]{Bragg2014} Bragg, A. D., \& Collins, L. R.\ 2014, New Journal of Physics, 16, 055014
\bibitem[Brauer et al.(2008a)]{Brauer2008a} Brauer, F., Henning, T., \& Dullemond, C.~P.\ 2008, \aap, 487, L1 
\bibitem[Brauer et al.(2008b)]{Brauer2008b} Brauer, F., Dullemond, C.~P., \& Henning, T.\ 2008, \aap, 480, 859 
\bibitem[Carrera et al.(2015)]{Carrera2015} Carrera, D., Johansen, A., \& Davies, M.~B.\ 2015, \aap, 579, A43 
\bibitem[Chiang \& Youdin(2010)]{Chiang2010} Chiang, E., \& Youdin, A.~N.\ 2010, Annual Review of Earth and Planetary Sciences, 38, 493 
\bibitem[Cuzzi \& Hogan(2003)]{Cuzzi2003} Cuzzi, J.~N., \& Hogan, R.~C.\ 2003, \icarus, 164, 127 
\bibitem[Davila \& Hunt(2001)]{Davila2001} Davila, J., \& Hunt, J. 2001, J. Fluid Mech., 440, 117
\bibitem[Donzis et al.(2008)]{Donzis2008} Donzis, D.~A., Yeung, P.~K., \& Sreenivasan, K.~R.\ 2008, Physics of Fluids, 20, 045108 
\bibitem[Donzis {et~al.}(2008)Donzis, Yeung, \& Sreenivasan]{Donzis_etal2008} Donzis, D.~A., Yeung, P.~K., \& Sreenivasan, K.~R. 2008, Phys. Fluids, 20, 045108
\bibitem[Falkovich \& Pumir(2007)]{Falkovich2007} Falkovich, G., \& Pumir, A.\ 2007, Journal of Atmospheric Sciences, 64, 4497 
\bibitem[Fessler et al.(1994)]{Fessler1994} Fessler, J.~R., Kulick, J.~D., \& Eaton, J.~K.\ 1994, Physics of Fluids, 6, 3742 
\bibitem[Garaud et al.(2013)]{Garaud2013} Garaud, P., Meru, F., Galvagni, M., \& Olczak, C.\ 2013, \apj, 764, 146 
\bibitem[Gunkelmann et al.(2016)]{Gunkelmann2016} Gunkelmann, N., Ringl, C., \& Urbassek, H.~M.\ 2016, \aap, 589, A30 
\bibitem[Gustavsson \& Mehlig(2011)]{Gustavsson2011} Gustavsson, K., \& Mehlig, B.\ 2011, \pre, 84, 045304 
\bibitem[Gustavsson et al.(2008)]{Gustavsson2008} Gustavsson, K., Mehlig, B., Wilkinson, M., \& Uski, V.\ 2008, Physical Review Letters, 101, 174503 
\bibitem[G{\"u}ttler et al.(2010)]{Gttler2010} G{\"u}ttler, C., Blum, J., Zsom, A., Ormel, C.~W., \& Dullemond, C.~P.\ 2010, \aap, 513, A56 
\bibitem[Hayashi(1981)]{Hayashi1981} Hayashi, C.\ 1981, Progress of Theoretical Physics Supplement, 70, 35 
\bibitem[Hubbard(2012)]{Hubbard2012} Hubbard, A.\ 2012, \mnras, 426, 784 
\bibitem[Ida \& Guillot(2016)]{Ida2016} Ida, S., \& Guillot, T.\ 2016, \aap, 596, L3 
\bibitem[Ireland et al.(2016a)]{Ireland2016a} Ireland, P.~J., Bragg, A.~D., \& Collins, L.~R.\ 2016, Journal of Fluid Mechanics, 796, 617 
\bibitem[Ireland et al.(2016b)]{Ireland2016b} Ireland, P.~J., Bragg, A.~D., \& Collins, L.~R.\ 2016, Journal of Fluid Mechanics, 796, 659 
\bibitem[Ishihara et al.(2007)]{Ishihara2007} Ishihara, T., Kaneda, Y., Yokokawa, M., Itakura, K., \& Uno, A.\ 2007, Journal of Fluid Mechanics,592, 335 
\bibitem[Ishihara et al.(2009)]{Ishihara2009} Ishihara, T., Gotoh, T., \& Kaneda, Y.\ 2009, Annual Review of Fluid Mechanics, 41, 165 
\bibitem[Ishihara {et~al.}(2015)Ishihara, Enohata, Morishita, Yokokawa, \&Ishii]{Ishihara2015} Ishihara, T., Enohata, K., Morishita, K., Yokokawa, M., \& Ishii, K. 2015, Parallel Computing Technologies, Vol. 9251 (Springer International Publishing), 522--527
\bibitem[Ishihara et al.(2016)]{Ishihara2016} Ishihara, T., Morishita, K., Yokokawa, M., Uno, A., \& Kaneda, Y.\ 2016, Physical Review Fluids, 1, 082403 \bibitem[Jimenez {et~al.}(1993)Jimenez, Wray, Saffman, \& Rogallo]{Jimenez1993} Jimenez, J., Wray, A., Saffman, P., \& Rogallo, R. 1993, J. Fluid Mech., 255
\bibitem[Johansen et al.(2009)]{Johansen2009} Johansen, A., Youdin, A., \& Mac Low, M.-M.\ 2009, \apjl, 704, L75 
\bibitem[Johansen et al.(2014)]{Johansen2014} Johansen, A., Blum, J., Tanaka, H., et al.\ 2014, Protostars and Planets VI, 547 
\bibitem[Kaneda et al.(2003)]{Kaneda2003} Kaneda, Y., Ishihara, T., Yokokawa, M., Itakura, K., \& Uno, A.\ 2003, Physics of Fluids, 15, L21 
\bibitem[Kataoka et al.(2013)]{Kataoka2013} Kataoka, A., Tanaka, H., Okuzumi, S., \& Wada, K.\ 2013, \aap, 557, L4 
\bibitem[Kerr(1985)]{Kerr1985} Kerr, R.~M. 1985, J.~Fluid Mech., 153, 31
\bibitem[Lissauer(1993)]{Lissauer1993} Lissauer, J.~J.\ 1993, \araa, 31, 129 
\bibitem[Markiewicz et al.(1991)]{Markiewicz1991} Markiewicz, W.~J., Mizuno, H., \& Voelk, H.~J.\ 1991, \aap, 242, 286 
\bibitem[Mattor {et~al.}(1995)Mattor, Williams, \& Hewett]{Mattor1995}Mattor, N., Williams, T.~J., \& Hewett, D. 1995, Parallel Computing, 21, 1769
\bibitem[Maxey(1987)]{Maxey1987} Maxey, M.~R.\ 1987, Physics of Fluids, 30, 1915 
\bibitem[Meru et al.(2013)]{Meru2013} Meru, F., Geretshauser, R.~J., Sch{\"a}fer, C., Speith, R., \& Kley, W.\ 2013, \mnras, 435, 2371 
\bibitem[Morishita {et~al.}(2015)Morishita, Yokokawa, Uno, Ishihara, \& Kaneda]{Morishita2015} Morishita, K., Yokokawa, M., Uno, A., Ishihara, T., \& Kaneda, Y. 2015, in Parallel CFD 2015
\bibitem[Okuzumi et al.(2009)]{Okuzumi2009} Okuzumi, S., Tanaka, H., \& Sakagami, M.-a.\ 2009, \apj, 707, 1247 
\bibitem[Okuzumi et al.(2012)]{Okuzumi2012} Okuzumi, S., Tanaka, H., Kobayashi, H., \& Wada, K.\ 2012, \apj, 752, 106 
\bibitem[Ormel \& Cuzzi(2007)]{Ormel2007} Ormel, C.~W., \& Cuzzi, J.~N.\ 2007, \aap, 466, 413 (OC07)
\bibitem[Pan \& Padoan(2013)]{Pan2013} Pan, L., \& Padoan, P.\ 2013, \apj, 776, 12 (PP13)
\bibitem[Pan \& Padoan(2014)]{Pan2014_IV} Pan, L., \& Padoan, P.\ 2014, \apj, 797, 101 
\bibitem[Pan \& Padoan(2015)]{Pan2015} Pan, L., \& Padoan, P.\ 2015, \apj, 812, 10 (PP15)
\bibitem[Pan et al.(2014a)]{Pan2014_II} Pan, L., Padoan, P., \& Scalo, J.\ 2014a, \apj, 791, 48 
\bibitem[Pan et al.(2014b)]{Pan2014_III} Pan, L., Padoan, P., \& Scalo, J.\ 2014b, \apj, 792, 69
\bibitem[Pan et al.(2011)]{Pan2011} Pan, L., Padoan, P., Scalo, J., Kritsuk, A.~G., \& Norman, M.~L.\ 2011, \apj, 740, 6 
\bibitem[Paszun \& Dominik(2009)]{Paszun2009} Paszun, D., \& Dominik, C.\ 2009, \aap, 507, 1023 
\bibitem[Saffman \& Turner(1995)]{Saffman1995} Saffman, P.~G., \& Turner, J.~S. 1995, J. Fluid Mech., 1, 16
\bibitem[Schumacher(2007)]{Schumacher2007} Schumacher, J. 2007, Europhysics Lett., 80, 54001
\bibitem[Seizinger \& Kley(2013)]{Seizinger2013} Seizinger, A., \& Kley, W.\ 2013, \aap, 551, A65 
\bibitem[Squires \& Eaton(1991)]{Squires1991} Squires, K.~D., \& Eaton, J.~K.\ 1991, Physics of Fluids A, 3, 1169 
\bibitem[Sundaram \& Collins(1997)]{Sundaram1997} Sundaram, S., \& Collins, L.~R.\ 1997, Journal of Fluid Mechanics, 335, 75 
\bibitem[V\"olk {et~al.}(1980)Volk, Jone, Morfill, \& Roeser]{Volk1980} Volk, H.~J., Jone, F.~C., Morfill, G.~E., \& Roeser, S. 1980, Astron. Astrophys., 85, 316
\bibitem[Vincent \& Meneguzzi(1991)]{Vincent1991} Vincent, A., \& Meneguzzi, M. 1991, J.~Fluid Mech., 225, 1 
\bibitem[Voelk et al.(1980)]{Voelk1980} Voelk, H.~J., Jones, F.~C., Morfill, G.~E., \& Roeser, S.\ 1980, \aap, 85, 316 
\bibitem[Wada et al.(2009)]{Wada2009} Wada, K., Tanaka, H., Suyama, T., Kimura, H., \& Yamamoto, T.\ 2009, \apj, 702, 1490
\bibitem[Wada et al.(2013)]{Wada2013} Wada, K., Tanaka, H., Okuzumi, S., et al.\ 2013, \aap, 559, A62 
\bibitem[Wang {et~al.}(2000)Wang, Wexler, \& Zhou]{Wang2000} Wang, L.-P., Wexler, A. S.~Q., \& Zhou, Y. 2000, J. Fluid Mech., 415
\bibitem[Watanabe \& Gotoh(2007)]{Watanabe2007} Watanabe, T., \& Gotoh, T. 2007, J. Fluid Mech., 590, 117
\bibitem[Weidenschilling(1977)]{Weidenschilling1977} Weidenschilling, S.~J.\ 1977, \mnras, 180, 57 
\bibitem[Wilkinson et al.(2006)]{Wilkinson2006} Wilkinson, M., Mehlig, B., \& Bezuglyy, V.\ 2006, Physical Review Letters, 97, 048501 
\bibitem[Windmark et al.(2012)]{Windmark2012} Windmark, F., Birnstiel, T., Ormel, C.~W., \& Dullemond, C.~P.\ 2012, \aap, 544, L16 
\bibitem[Yamazaki et al.(2002)]{Yamazaki2002} Yamazaki, Y., Ishihara, T., \& Kaneda, Y.\ 2002, Journal of the Physical Society of Japan, 71, 777
\bibitem[Yang et al.(2017)]{Yang2017} Yang, C.-C., Johansen, A., \& Carrera, D.\ 2017, \aap, 606, A80 
\bibitem[Yeung \& Pope(1988)]{Yeung1988} Yeung, P.~K., \& Pope, S.~B. 1988, Journal of Computational Physics, 79, 373
\bibitem[Yeung {et~al.}(2015)Yeung, Zhai, \& Sreenivasan]{Yeung2015} Yeung, P.~K., Zhai, X.~M., \& Sreenivasan, K.~R. 2015, Proc. Natl. Acad. Sci. USA, 112, 12633
\bibitem[Yokokawa {et~al.}(2002)Yokokawa, Itakura, Uno, Ishihara, \& Kaneda]{Yokokawa2002}Yokokawa, M., Itakura, K., Uno, A., Ishihara, T., \& Kaneda, Y. 2002, in Proceedings of the IEEE/ACM SC2002 Conference, 50
\bibitem[Youdin \& Goodman(2005)]{YG05} Youdin, A.~N., \& Goodman, J.\ 2005, \apj, 620, 459 
\bibitem[Zsom et al.(2010)]{Zsom2010} Zsom, A., Ormel, C.~W., G{\"u}ttler, C., Blum, J., \& Dullemond, C.~P.\ 2010, \aap, 513, A57 
\bibitem[Zsom et al.(2011)]{Zsom2011} Zsom, A., Ormel, C.~W., Dullemond, C.~P., \& Henning, T.\ 2011, \aap, 534, A73 

\end{thebibliography}

\end{document}